\documentclass[%
superscriptaddress,
amsmath,amssymb,
aps,
]{revtex4-2}

\usepackage{graphicx}
\usepackage{dcolumn}
\usepackage{bm}
\usepackage{hyperref}

\usepackage{braket}
\usepackage{color}
\usepackage{lineno}
\usepackage{pdfpages}
\usepackage{etoolbox} 

\makeatletter
\patchcmd{\@outputpage@head}{\@ifx{\LS@rot\@undefined}{}{\LS@rot}}{}{}{}
\makeatother

\begin{document}


# Vortex dynamics in the two-dimensional BCS-BEC crossover

Max Heyl[1,2]†, Kyosuke Adachi[3,4]†, Yuki M. Itahashi[1], Yuji Nakagawa[1], Yuichi Kasahara[5], Emil J. W. List-Kratochvil[2,6], Yusuke Kato[7], and Yoshihiro Iwasa[1,8]*

[1]Quantum-Phase Electronics Center and Department of Applied Physics, University of Tokyo, Hongo 7-3-1, Bunkyo-ku, Tokyo 113-8656, Japan.

[2]Department of Chemistry, Department of Physics & IRIS Adlershof, Humboldt-Universität zu Berlin, Zum Großen Windkanal 2, 12489 Berlin, Germany.

[3]Nonequilibrium Physics of Living Matter RIKEN Hakubi Research Team, RIKEN Center for Biosystems Dynamics Research, 2-2-3 Minatojima-minamimachi, Chuo-ku, Kobe 650-0047, Japan.

[4]RIKEN Interdisciplinary Theoretical and Mathematical Sciences Program, 2-1 Hirosawa, Wako 351-0198, Japan.

[5]Department of Physics, Kyoto University, Kitashirakawa Oiwakecho, Sakyo-ku, Kyoto 606-8502, Japan.

[6]Helmholtz-Zentrum für Materialien und Energie GmbH, Hahn-Meitner-Platz 1, 14109 Berlin, Germany.

[7]Department of Basic Science, University of Tokyo, Meguro-ku, Tokyo 153-8902, Japan.

[8]RIKEN, Center for Emergent Matter Science, Hirosawa 2-1, Wako 351-0198, Japan.

*E-mail: iwasa@ap.t.u-tokyo.ac.jp
†These authors contributed equally to this work.

**The Bardeen-Cooper-Schrieffer (BCS) condensation and Bose-Einstein condensation (BEC) are the two limiting ground states of paired Fermion systems, and the crossover between these two limits has been a source of excitement for both fields of high temperature superconductivity and cold atom superfluidity. For superconductors, ultra-low doping systems like graphene and $Li_xZrNCl$ successfully approached the crossover starting from the BCS-side. These superconductors offer new opportunities to clarify the nature of charged-particles transport towards the BEC regime. Here we report the study of vortex dynamics within the crossover using their Hall effect as a probe in $Li_x$ZrNCl. We observed a systematic enhancement of the Hall angle towards the BCS-BEC crossover, which was qualitatively reproduced by the phenomenological time-dependent Ginzburg-Landau (TDGL) theory. $Li_x$ZrNCl exhibits a band structure free from various electronic instabilities, allowing us to achieve a comprehensive understanding of the vortex Hall effect and thereby propose a global picture of vortex dynamics within the crossover. These results demonstrate that gate-controlled superconductors are ideal platforms towards investigations of unexplored properties in BEC superconductors.**

**Introduction**

The crossover between the two limiting ground states of Fermion systems - the BCS and BEC state - has attracted continuous interest both theoretically and experimentally from the communities of ultracold atomic gases and superconductors[1-3]. The first experimental realization of the BCS-BEC crossover was achieved in ultracold atomic gases[4] starting from the BEC side, while the approach with superconductors from the BCS side[5] has become active since recent discoveries of suitable materials including FeSe[6], twisted graphene[7,8], and $Li_x$ZrNCl[9]. The two BCS-BEC crossover systems, twisted trilayer graphene and $Li_x$ZrNCl, are highly two-dimensional (2D), and the carrier density can be controlled by a gate voltage. The

2

tunable carrier density is highly advantageous to observe how the system evolves from the BCS- to the BEC-limit. In fact, both systems approached the crossover regime by reducing the carrier density and reached $T_c/T_F \sim 1/8$, where $T_c$ and $T_F$ are the critical temperature and the Fermi temperature, respectively. Since $T_c/T_F = 1/8$ is the upper limit for 2D systems[10,11] in the crossover, it confirms the successful approach of the 2D BCS-BEC crossover. By approaching the crossover towards the BEC-limit, the coupling strength described as the ratio of superconducting gap $\Delta$ to Fermi energy $E_F$ increased[9]. One of the intriguing phenomena related to the enhancement of $\Delta/E_F$ is the dynamics of superconducting vortices. Figures 1a and 1b show a comparison of the energy levels inside the vortices for the BEC and BCS limit. Inside the vortex core, the quasiparticle states are confined and quantized. According to the Caroli-de Gennes-Matricon picture[13], the energy level spacing is in the order of $\Delta^2/E_F$. In the BCS limit, $\Delta^2/E_F$ is so small that the energy spectrum is almost continuous (Fig. 1b), and thus the quasiparticles in the core are easily scattered, resulting in energy dissipative vortex motion. In the BEC limit, on the other hand, the energy level spacing is large enough to form a single quantized level[14] (Fig. 1a). This large level spacing renders the vortex motion dissipationless since the quasiparticles in the core cannot be scattered. The above difference results in contrasting vortex motions: the vortex in the BCS regime moves perpendicularly to the net supercurrent (Fig. 1b, arrows), while the vortex in the BEC regime moves parallel to the superfluid velocity, i.e., anti-parallel to the net supercurrent (Fig. 1a, arrows). In the BEC regime, the motion of vortices antiparallel (parallel) to the background current density $\boldsymbol{J}$ (superfluid velocity) can be understood as follows: The vortex motion in the presence of the magnetic field $\mathbf{B}$ induces the macroscopic electric field $\boldsymbol{E} = \boldsymbol{B} \times \boldsymbol{v}_v$, where $\boldsymbol{v}_v$ is the vortex velocity. In the absence of dissipation in the steady state, $\boldsymbol{E} = \boldsymbol{B} \times \boldsymbol{v}_v$ should be perpendicular to $\boldsymbol{J}$ to avoid acceleration of the supercurrent. Thus $\boldsymbol{J} \parallel \boldsymbol{v}_v$ follows. Hence, the dissipationless nature of the vortex core in the BEC regime (Fig. 1a) brings about vortex flow parallel or antiparallel to the supercurrent flow. This vortex flow parallel to the supercurrent

has never been observed experimentally, but in the case of superfluidity of charge neutral bosons of $^4$He, the parallel vortex flow was nicely demonstrated in the movie of Ref. 15 (for more details on the analogy between single vortex dynamics and charged particle dynamics see Supplementary Sec. 1).

In superfluid $^4$He, the vortex motion is parallel to the superfluid velocity at temperatures much lower than $T_c$[15], whereas observation of the vortex flow anti-parallel to the current in superconductivity has been hindered by the lack of suitable materials. The recently established gate-controlled BCS-BEC crossover is a promising candidate system to investigate the anomalous vortex flow ever closer to the BEC regime. To probe the anomalous vortex motion, we focus on the Hall effect of vortices and its evolution with the carrier density.

The vortex Hall effect (VHE) is a unique transport phenomenon of superconductors, where the collective dynamics of vortices and associated magnetic fluxes produces the Hall voltage[16,17]. The VHE has been recognized since the middle of the last century in several metallic superconductors[18] and was most intensively investigated in high $T_c$ cuprates[16,19–23]. Though the potential importance of the BCS-BEC crossover has been discussed for cuprates[1,24], its impact on the Hall effect[25] and vortex dynamics remains elusive due to the orders competing with superconductivity such as antiferromagnetism and charge density wave[26]. In particular, it is still unknown how the temperature-dependent sign change of the VHE, which is called Hall anomaly[16], is related to the BCS-BEC crossover. In the gate-controlled Li$_x$ZrNCl system, where no competing orders have been observed, intact effects of the BCS-BEC crossover on the VHE and vortex dynamics are expected to be unveiled by transport measurements combined with systematic control of $\Delta/E_F$. Even comparing with highly controllable ultracold atomic gases, where transport measurements are developing[27,28] but still challenging[29], Li$_x$ZrNCl is a suitable system to systematically study the vortex dynamics in the BCS-BEC crossover.

Here we report the evolution of the VHE in the BCS-BEC crossover regime via gate-controlled superconductivity of Li$_x$ZrNCl. We experimentally found that the Hall angle, related to vortex motion, increases with decreasing Li content $x$ or electron density towards the crossover regime. Using the time-dependent Ginzburg-Landau (TDGL) model, we successfully explained the temperature dependence of the transverse resistivity and the $x$ dependence of the Hall angle in a qualitative manner. The combination of experiments and theory enables us to present a comprehensive picture of vortex dynamics along the BCS-BEC crossover. Leaving the experimental region in the crossover, the expected vortex dynamics in the BEC limit were conjectured, allowing to present a full view on the evolution of vortex dynamics. Therefore, this work shows that the Li$_x$ZrNCl system allows for an unclouded view on the BCS-BEC crossover rendering it an ideal testbed to benchmark theories on vortex dynamics in the crossover.

**Results**

Experimental observation of the vortex Hall effect in Li$_x$ZrNCl

Figures 1c and 1d depict side and top view, respectively, of the crystal structure of the Li-intercalated ZrNCl system (Li$_x$ZrNCl). The host ZrNCl is a van der Waals layered material with a double honeycomb lattice composed of Zr and N, forming a band insulator. Once Li is intercalated in the van der Waals gap, one electron per Li is introduced to the ZrN conduction layer, and the system exhibits superconductivity[12]. The Li concentration $x$ corresponds to the carrier density. Figure 1e displays a schematic band structure of Li$_x$ZrNCl. Electrons are introduced into the parabolic conduction bands, which are located at the K and K' points at the corners of the hexagonal Brillouin zone. The electronic phase diagram in the temperature $T$ and electron density $x$ plane is shown in Fig. 1f. The pseudo-gap state appears at rather high temperature $T^*$, and the superconducting transition temperature $T_c$ exhibits a maximum at $x \sim$ 0.01, below which $T_c$ is well scaled as $T_c/T_F \sim 1/8$. Furthermore, the ratio of superconducting

gap $\Delta$ to Fermi energy $E_F$ increases with decreasing $x$[9]. The determination of $\Delta$ and $E_F$, which are not specific to superconductor and cold atom systems, enables us to construct a unified experimental phase diagram of the BCS-BEC crossover (Supplementary Sec. 2 and Supplementary Fig. 1). This encourages us to investigate the nature of BEC superconductivity by extrapolating the trend from the BCS side.

To navigate the single-crystal ZrNCl within the BCS-BEC crossover, an intercalation-only device was employed as previously reported[9,30]. Several doping levels were achieved (Fig. 1f, Supplementary Sec. 3, and Supplementary Fig. 2). The $T_c$ data obtained in the present experiment (dark blue) agrees well with the previous results. This phase diagram allows us to correlate observed transport phenomena with the position in the crossover.

A detailed look at the transport properties is given with $x = 0.0040$, $0.010$, and $0.47$ in Fig. 2, which shows the longitudinal and transverse resistivities versus temperature at varying out-of-plane magnetic fields. For the lowest doping, $x = 0.0040$ (Fig. 2a), superconductivity ($T_c = 16.8$ K) persists even at elevated magnetic fields, and the upper critical field $B_{c2}$ reaches 5.9 T (Supplementary Sec. 4 and Supplementary Fig. 3). No quantum metallic states are visible unlike the electrostatically induced monolayer or bilayer superconductivity[31]. This discrepancy is possibly because the present system is multilayered, similarly to the bulk cuprates. The transverse resistivity $\rho_{yx}$ (bottom) exhibits anomalous behavior. For $T > T_c$, the Hall signal is negative ($\rho_{yx} < 0$), as expected for the n-type transport in electron-doped Li$_x$ZrNCl. For $T \sim T_c$, however, the transverse resistivity starts to change its sign to $\rho_{yx} > 0$ and forms a peak with reducing $T$, which is reminiscent of the Hall anomaly[16]. For medium doping, $x = 0.010$ (Fig. 2b), superconductivity ($T_c = 15.9$ K) is less persistent at elevated fields ($B_{c2} = 5.0$ T), and the Hall anomaly is observed though the magnitude is reduced. For high doping, $x = 0.47$ (Fig. 2c), the system shows a sharp transition ($T_c = 11.4$ K), a much lower critical field ($B_{c2} = 0.8$ T), and an extremely small VHE. It is important to note that the "cleanness", i.e., the ratio of the mean free path over the coherence length has been observed

to play a crucial role in the appearance of the Hall anomaly as discussed previously[16]. The doping dependence of the cleanness is shown in Supplementary Fig. 4, to exclude that the observed enhancement of the Hall anomaly versus doping is largely due to a modulation of the cleanness (see Supplementary Sec. 5 and Supplementary Table 1 for experimentally obtained parameters, including the mean free path and the coherence length).

The doping dependence of the VHE can be visualized by mapping the Hall coefficient $R_H$ (= $\rho_{yx}/B$) at different doping levels as a function of temperature and field (Fig. 3a). Blue and red areas correspond to the negative and positive $R_H$, respectively. The red color highlights the enhanced VHE, which appears in the low-field regime around the $B_{c2}$ curve, where vortices are highly mobile. Here, the $B_{c2}$ curve for each carrier density is determined by the half resistance temperature. These features strongly support that the observed Hall anomalies indeed reflect the VHE. At $x = 0.47$, the VHE almost vanishes. Overall, it is evident that the magnitude and area of the VHE are increasing with decreasing doping. The maximum Hall angle $\Theta_H$, defined at the dark red circles in each phase diagram (Fig. 3a), is summarized as a function of $x$ in Fig. 3b. Though the temperature and field, at which $\Theta_H$ is defined (Fig. 3a), vary from sample to sample, $\Theta_H$ characterizes the strength of the VHE at each doping level. The observed $x$ dependence of $\Theta_H$ unambiguously implies that the VHE is enhanced by reducing the carrier density, i.e., approaching the BCS-BEC crossover from the BCS side.

Theoretical modelling of the experimentally observed vortex Hall effect

To elucidate the evolution of vortex dynamics towards the BCS-BEC crossover, observed as the enhanced VHE, we calculated $\rho_{xx}$ and $\rho_{yx}$ using the 2D TDGL model (equation (1) in Methods) with the gauge invariance[32,33]. Since the sign change of $\Theta_H$ was observed both above and below the $B_{c2}$ curve (Fig. 3a), we focused on the dynamics in the

vortex liquid state[34], where many interacting vortices are fluctuating in time and space[17]. Using a few phenomenological parameters and the $E_F$ dependence of the gap-opening temperature $T^*$ (Supplementary Fig. 5), which has been measured previously[9], we obtained the longitudinal and transverse resistivities near the $B_{c2}$ curve (Methods and Supplementary Sec. 6-9).

Figures 4a and 4b show the obtained temperature dependence of $\rho_{xx}$ (top) and $\rho_{yx}$ (bottom) for $x$ = 0.0040 and 0.47, respectively. As shown in these figures, we find a qualitative agreement with the experimentally observed temperature and field dependence of $\rho_{xx}$ and $\rho_{yx}$, especially the sign change and positive peak of $\rho_{yx}$ for $x$ = 0.0040 (Fig. 2a). This agreement indicates that the peak of $\rho_{yx}$ occurs in the vortex liquid state, without considering pinning effects. The disappearance of the clear peak in $\rho_{yx}$ for increased $x$ to 0.47 (Fig. 4b) also agrees well with the experiment (Fig. 2c). We have confirmed that these agreements can be obtained even if we change a parameter that is not determined from the experiments (Supplementary Figs. 6 and 7). There are still some discrepancies between experiment and theory. For instance, in the low-temperature high-field region for $x$ = 0.0040, however, the VHE is suppressed in the experiment (Fig. 2a) but still visible in the theory (Fig. 4a). The absence of $\rho_{yx}$ despite the finite $\rho_{xx}$ is reminiscent of the quantum vortex liquid state in 2D superconductors[35]. In this state, the vortex is in the liquid state by quantum fluctuation rather than thermal fluctuation[36,37]. This quantum vortex liquid state is left to be confirmed in future studies. In particular, the development of a TDGL model including quantum fluctuations is highly anticipated.

Figure 4c compares the theoretically obtained Hall angle (Methods and Supplementary Sec. 7) and the experimentally observed $\Theta_H$ (Fig. 3b), which show similar trends of enhancement towards the BCS-BEC crossover. Strictly speaking, the experimental $\Theta_H$ was measured at the peak position of $R_H$ (Fig. 3a), while the theoretical Hall angle was

determined at low temperatures where the vortex contribution to the conductivity is dominant (Supplementary Sec. 7). Nevertheless, the similar trends in the theory and experiment suggest that the TDGL model successfully captures the evolution of the vortex dynamics and the resulting VHE towards the BCS-BEC crossover. The quantitative difference between the theory and experiment might be attributed to theoretical overestimation of the fluctuation contribution at high fields (see Supplementary Sec. 9).

The key to the above agreement is that the sign of the Hall angle in the vortex state is opposite to the sign of the $E_F$ derivative of the mean-field critical temperature in the TDGL model[33] (equation (2) in Methods). In the BCS-BEC crossover, the mean-field critical temperature, which represents the pairing temperature[2,3,24], should correspond to the gap-opening temperature $T^*$ rather than the observed critical temperature $T_c$. Hence, the sign of the vortex Hall angle, and thereby the occurrence of the Hall anomaly, is determined from that of $dT^*/dE_F$, which generally depends on the microscopic details of the system. Since $dT^*/dE_F < 0$ in Li$_x$ZrNCl (Supplementary Fig. 5), the Hall anomaly observed as the sign change of $\Theta_H$ from negative to positive upon cooling towards the vortex state was theoretically reproduced. This situation ($dT^*/dE_F < 0$) can be microscopically derived from a tight binding model with a finite-range attractive interaction[38]. The simple material nature of Li$_x$ZrNCl seems important for the applicability of the TDGL model because, in many cuprate superconductors, the simple TDGL model is not applied possibly due to several complex features including the $d$-wave superconductivity[20] or competing orders such as antiferromagnetism and charge density wave[26].

**Discussion**

The simple material nature of Li$_x$ZrNCl motivated us to conjecture the evolution of vortex dynamics beyond the experimentally established region in the crossover (Fig. 4c, Fig. 4d light blue curve) towards the BEC limit (Fig. 4d green curve). This way, Figure 4d displays the

9

expected global evolution of the Hall angle $\Theta_\mathrm{H}^\mathrm{vortex}$ in the vortex state with changing the doping level $x$ from the BEC to the BCS limit, accompanied with schematics of vortex motions in the two limits. Here we assumed that the vortex state stays liquid over all regions. In the BCS limit, the vortex flow (thin red arrow) is perpendicular to the supercurrent (blue arrow), indicating $\Theta_\mathrm{H}^\mathrm{vortex} = 0$. The experimentally observed region (light blue line) spans from the BCS to the crossover regime, where the tilting of the vortex motion increases $\Theta_\mathrm{H}$ as $x$ is reduced. This behavior is supported by the theoretical expression of $\Theta_\mathrm{H}^\mathrm{vortex}$, i.e., $\tan\Theta_\mathrm{H}^\mathrm{vortex} = \lambda/\gamma \sim O(T^*/E_\mathrm{F})$ (equation (2) in Methods and Supplementary Sec. 7). Here, the coefficients $\gamma$ and $\lambda$ represent typical times for dissipative damping and non-dissipative propagation, respectively. In other words, this equation relates the vortex Hall angle to the relative importance of dissipation ($\gamma/\lambda$): the vortex Hall angle is becoming smaller as the dissipation is growing larger. Thus, in the BCS region, where $T^*/E_\mathrm{F} \simeq 0$, the dissipation is dominant ($\lambda/\gamma \simeq 0$) and the vortex Hall angle is almost zero. Furthermore, this relation suggests that $|\Theta_\mathrm{H}^\mathrm{vortex}|$ should increase as $T^*/E_\mathrm{F}$ increases towards the crossover region[9]. In the BEC limit as the low-doping limit, on the other hand, vortices will move anti-parallel to the supercurrent due to the dissipationless vortex core (Fig. 1a), leading to a large negative $\Theta_\mathrm{H}^\mathrm{vortex}$. This picture in the dilute limit is also supported by the TDGL model derived from the dilute Fermi gas model with attractive interaction (Supplementary Sec. 10), which gives vanishing dissipation[24] ($\gamma = 0$ in equation (1) in Methods) and accordingly $\Theta_\mathrm{H}^\mathrm{vortex} = -\pi/2$ (green line in Fig. 4d). Thus, if one can further reduce $x$ from the experimentally observed region, $\Theta_\mathrm{H}^\mathrm{vortex}$ is expected to form a positive peak, change its sign, and eventually approach $-\pi/2$ (black dotted line in Fig. 4d). Overall, through the BCS-BEC crossover in Li$_x$ZrNCl, the vortex dynamics will non-monotonically evolve from dissipative motion perpendicular to the supercurrent into dissipationless motion anti-parallel to the supercurrent.

Figure 4d also provides a comprehensive view of the Hall anomaly throughout the BCS-BEC crossover, corroborated by the TDGL theory. Since $\Theta_\text{H}$ is negative (i.e., $\sigma_{yx} > 0$, see Supplementary Table 2) in the high-temperature normal state of Li$_x$ZrNCl, the Hall anomaly, i.e., the sign reversal of $\Theta_\text{H}$ on lowering the temperature (Fig. 2 and Fig 4a), is only expected in the intermediate BCS-BEC crossover regime where $\Theta_\text{H}^\text{vortex}$ is positive due to the negative sign of d$T^*$/d$E_\text{F}$. (Fig. 4d).

In conclusion, we have systematically investigated the vortex Hall effect in the 2D BCS-BEC crossover using the recently established gate-controlled Li$_x$ZrNCl system on the experimental side. We experimentally observed the enhancement of the Hall angle with decreasing doping within the crossover. Combined with the TDGL theory, we successfully reproduced this behaviour qualitatively, and therefore present the experimentally supported evolution of vortex dynamics within the crossover. This model also allowed to conjecture the expected vortex dynamics beyond the experimental region, towards the BEC limit, unlocking a comprehensive picture of vortex dynamics throughout and beyond the BCS-BEC crossover. The simple band structure of Li$_x$ZrNCl with absence of competing orders should be crucial to establish a clear understanding of the VHE in many superconductors including cuprates, which has been left unsolved for a long time. The presented density-controlled BCS-BEC crossover has the potential to serve as an essential platform for material exploration and investigations of the unexplored physics of superconductivity in the BEC limit, as showcased herein with the open question of vortex dynamics in the BEC limit left to be tackled experimentally.

# Methods

**Device fabrication**

Bulk ZrNCl, prepared by a chemical vapor transport method[39], was exfoliated onto $SiO_2$/Si substrates using the scotch tape technique. The obtained single-crystalline thin flakes exhibited a thickness of ca. 10 – 40 nm, as determined by atomic force microscopy in tapping mode. Using electron-beam lithography (EBL) with polymethylmethacrylate (PMMA) as positive resist, Au (90 nm)/Cr (7 nm) electrodes were patterned onto a selected thin-flake in a Hall bar setup including a co-planar gate electrode in proximity. In a second EBL step only the edges of the ZrNCl thin flakes and the gate pad were developed, leaving the Hall bar electrodes and the channel region covered with PMMA, to allow for intercalation-only operation of the device[9,30]. For the electrolyte, $LiClO_4$ (Sigma Aldrich) was dissolved in polyethylene glycol (PEG, $M_w$ = 600, Wako) at a Li:O (PEG) ratio of 1:20 and stored at 80°C under vacuum. A drop of electrolyte was placed onto the device to cover both the exposed flake edges and the gate pad. A small cover glass slip was placed on top to evenly spread the electrolyte. The device was transferred into a Quantum Design Physical Property Measurement System (PPMS) equipped with a rotator probe and high vacuum was applied ($<10^{-4}$Torr) at 330 K for at least 1 h before measurement.

**Transport measurements**

The temperature-dependent resistance of the device at varying magnetic fields was measured in standard four-probe geometry using the PPMS combined with lock-in amplifiers (Stanford Research Systems Model SR830 DSP and Signal Recovery Model 5210) to measure current and voltage. A Keithley 2400 SMU was used to apply the gate voltage at 330 K and high vacuum. The temperature was lowered to 150 K to freeze the PEG-based electrolyte ($T_m$= 288 K), the chamber was purged with He and the Hall coefficients were measured.

**TDGL model**

Assuming that $T^*$ corresponds to the mean-field critical temperature, we used the phenomenological two-dimensional (2D) time-dependent Ginzburg-Landau (TDGL) model:

$$(\gamma + i\lambda)\frac{\partial}{\partial t}\Delta(\mathbf{r},t) = -\left[\frac{T-T^*}{T^*} + b|\Delta(\mathbf{r},t)|^2 - \xi^2\left(\nabla + i\frac{2\pi}{\phi_0}\mathbf{A}(\mathbf{r})\right)^2\right]\Delta(\mathbf{r},t), \quad (1)$$

where $b$ is a phenomenological parameter, $\phi_0$ is the flux quantum, $\xi$ $[= (\phi_0/2\pi B_{c2})^{1/2}]$ is the coherence length, and $\Delta(\mathbf{r}, t)$ is the superconducting order parameter varying in space and time. For the left-hand side, we simply used the relaxation time derived for the BCS regime[40], $\gamma = \pi/8T^*$, and took $\lambda = -(1/2T^*)\, \partial T^*/\partial E_F$, according to the gauge invariance[32,33]. For the expression of $\lambda$, we used the mean-field critical temperature $T^*$ instead of the observed transition temperature $T_c$ since the derivation in Aronov et al.[33] is based only on the linear terms in the TDGL model (1). We also set the chemical potential to $E_F$ in the expression of $\lambda$ since the chemical potential presumably takes a value similar to $E_F$ in the experimentally accessible BCS-BEC crossover region; considering the 2D Fermi gas model[10] as a reference, we see $\mu \sim 0.96\, E_F$ even when the superconducting gap at the BKT transition temperature is large as $\Delta = 0.4\, E_F$, which is in the same order as the observed value in $Li_xZrNCl$[9]. We set the vector potential as $\mathbf{A}(\mathbf{r}) = Bx\hat{y}$ in the Landau gauge, which is fixed so that the scalar potential is zero.

**Calculation of longitudinal and transverse conductivities**

Based on the TDGL model, after renormalizing the transition temperature from $T^*$ to $T_c$ (Supplementary Sec. 6), we applied the linear response theory within the Hartree approximation to obtain the temperature and field dependence of the longitudinal and transverse conductivities ($\sigma^V_{xx}$ and $\sigma^V_{yx}$) for the vortex liquid state[34] (Supplementary Sec. 7).

To calculate the resistivities ($\rho_{xx}$ and $\rho_{yx}$), we introduced the normal-state longitudinal and transverse conductivities ($\sigma^N_{xx}$ and $\sigma^N_{yx}$), which were obtained from the experimental data (Supplementary Table 2). For a phenomenological parameter $\beta$ ($\propto b$), which represents the fluctuation interaction strength, we typically took $\beta = 10^{-4} - 10^{-3}$, a slight change of which did not affect the qualitative behavior of $\rho_{xx}$ or $\rho_{yx}$ (Supplementary Sec. 9). In Figs. 4a and 4b, we show the obtained $\rho_{xx} = (\sigma^N_{xx} + \sigma^V_{xx})/[(\sigma^N_{xx} + \sigma^V_{xx})^2 + (\sigma^N_{yx} + \sigma^V_{yx})^2]$ and $\rho_{yx} = -(\sigma^N_{yx} + \sigma^V_{yx})/[(\sigma^N_{xx} + \sigma^V_{xx})^2 + (\sigma^N_{yx} + \sigma^V_{yx})^2]$. The Hall angle for low enough temperatures can be obtained as (Supplementary Sec. 7)

$$\Theta_H^{\text{vortex}} = -\arctan\left(\frac{4}{\pi}\frac{\partial T^*}{\partial E_F}\right), \quad (2)$$

which is plotted in Fig. 4c.

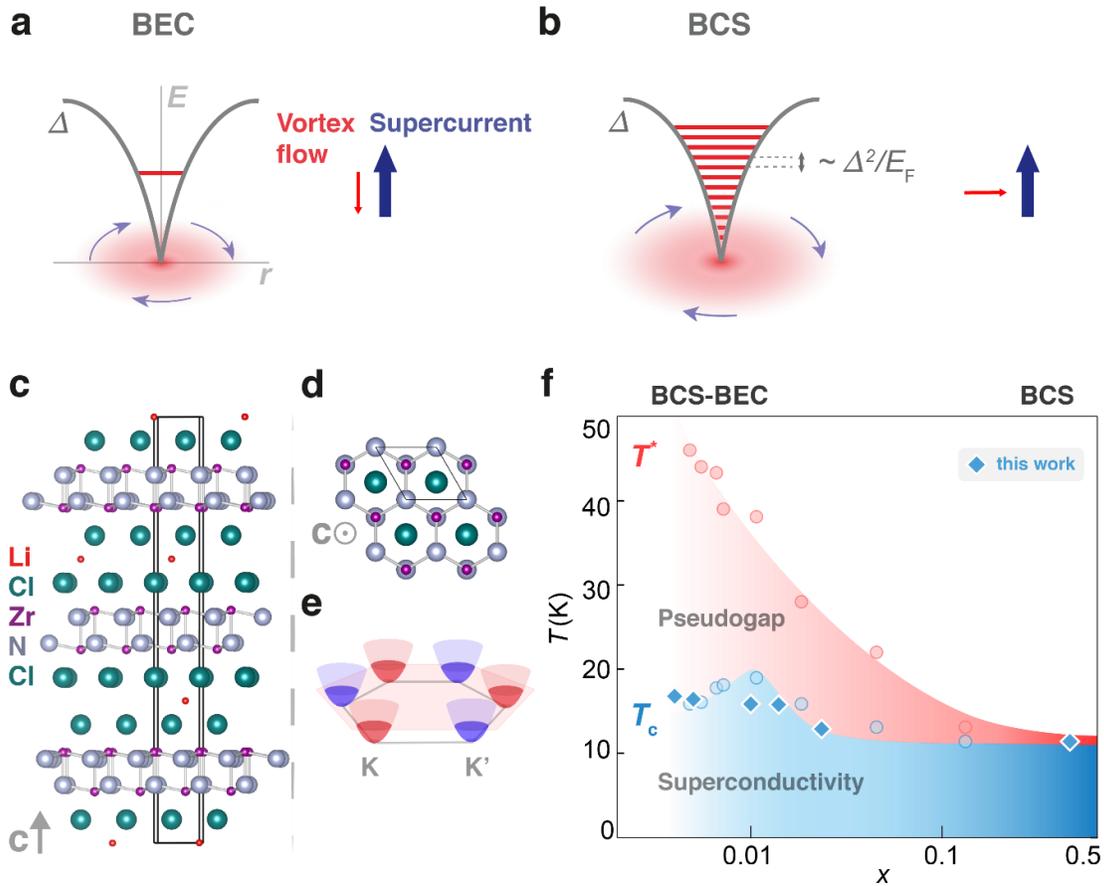

**Fig. 1 | Properties of superconducting Li$_x$ZrNCl.**

**a,b**) Caroli-de Gennes-Matricone (CdGM) schematic of the quantized state(s) in the vortex core in the BEC (**a**) and BCS limit (**b**) are depicted. The directions of supercurrent and vortex flow are indicated. In the BEC limit, the vortex motion is dissipationless and the vortex flow is nearly anti-parallel to the supercurrent, while in the BCS limit the vortex motion is nearly perpendicular to the supercurrent. Side (**c**) and top (**d**) view of the Li$_x$ZrNCl crystal structure. The solid black lines represent the hexagonal unit cell. **e**, Schematic of the simple parabolic conduction bands of Li$_x$ZrNCl, located at the corners of the hexagonal Brillouin zone (K and K' points). **f**, The BCS-BEC crossover phase diagram of Li$_x$ZrNCl[9] with data points from this work superimposed. $T_c$ was determined by the temperature at which the resistance is half the normal state value at 30 K, which is close enough to the Berezinskii-Kosterlitz-Thouless

transition for 2D superconductors[9]. The gap-opening temperature $T^*$ was determined in the previous study[9], and the regime between $T_c$ and $T^*$ forms the pseudo-gap state.

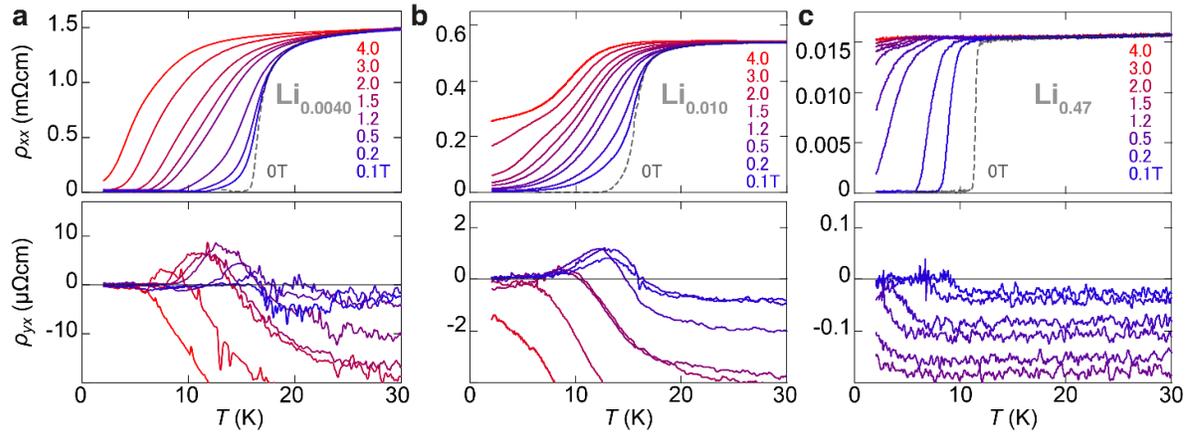

**Fig. 2 | Transport properties at different doping levels *x*.**

Temperature dependence of longitudinal $\rho_{xx}$ (top) and transverse $\rho_{yx}$ (bottom) resistivity at varying out-of-plane fields for $x$ = 0.0040 (**a**), $x$ = 0.010 (**b**), and $x$ = 0.47 (**c**). The vortex Hall effect with sign reversal is clearly observed in (**a**) and (**b**), whereas it is almost indiscernible in (**c**). The magnitude of $\rho_{yx}$ decreases with increasing $x$.

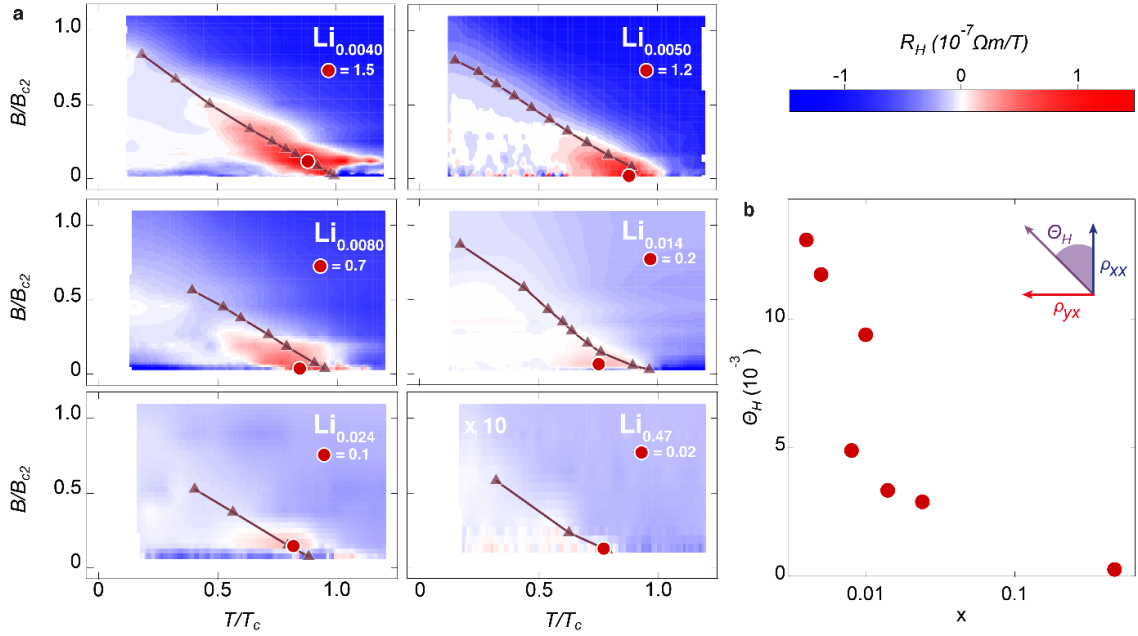

**Fig. 3 | Mapping the vortex Hall effect at different doping levels.**

**a**, Color mapping of the Hall coefficient $R_H$ on the $B$-$T$ phase diagram for different doping levels. Horizontal and vertical axes are temperature normalized by $T_c$ and magnetic field normalized by upper critical field $B_{c2}$ at zero temperature, respectively. The VHE is visible in the red regions with a sign change of $R_H$. The areas of VHE are increasing with decreased doping. At the highest doping ($x$ = 0.47), the VHE is not visible anymore. Lines and markers show the $T_c$ ($B_{c2}$) superconductivity boundary determined by the half resistance. Red circles mark the maximum $R_H$ values. **b**, Doping dependence of Hall angle $\Theta_H$. With decreasing doping, $\Theta_H$ is enhanced monotonically. In the high doping BCS regime, $\Theta_H$ almost vanishes.

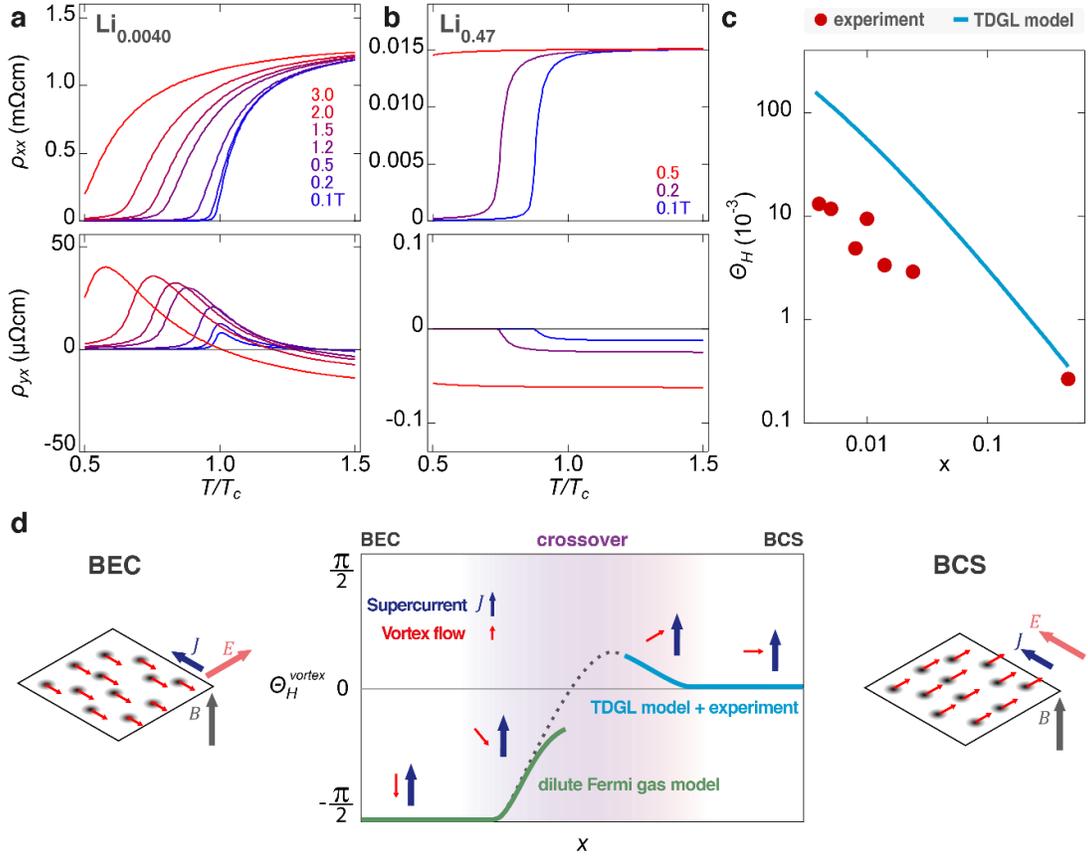

**Fig. 4 | Theoretical simulations of VHE by the TDGL framework.**

Temperature dependence of the longitudinal resistivity $\rho_{xx}$ (top) and the transverse resistivity $\rho_{yx}$ (bottom) at varying out-of-plane fields for the selected doping levels $x$ = 0.0040 (a) and 0.47 (b). c, Comparison of the theoretical and experimental doping dependence of the Hall angle $\Theta_H$. The experimental data are taken from Fig. 3, whereas the theoretical values are obtained for lower temperatures, where the normal-state contribution is negligible (Supplementary Sec. 7). d, Evolution of the vortex Hall angle $\Theta_H^{vortex}$ across the BCS-BEC crossover conjectured from the present study. The light blue curve is obtained from the experiment and TDGL model, while the light green curve is obtained theoretically using the diluted Fermi gas model. The dotted line is the interpolation between these two regimes. The blue and red arrows in the inset show the supercurrent and vortex flow, respectively.

21

Schematics in both sides illustrate these relative motions for a set of vortices at the BEC (left) and BCS (right) limits. Here J, E and B are the supercurrent flow, electric field induced by vortex flow, and external magnetic field, respectively. The present experiment is made around the border of the BCS and crossover regimes. It is important to note, the green curve in the BEC regime is a theoretical expectation via the dilute Fermi gas model and is beyond the experimentally supported region of the crossover and BCS regime.


\renewcommand{\thetable}{\arabic{table}}
\renewcommand{\tablename}{Supplementary Table}
\renewcommand{\figurename}{Supplementary Figure}
\renewcommand{\theequation}{S\arabic{equation}}

\title{Supplementary information:\\
Vortex dynamics in the two-dimensional BCS-BEC crossover}

\author{Max Heyl$^{1,2\dagger}$, Kyosuke Adachi$^{3,4\dagger}$, Yuki M. Itahashi$^{1}$, Yuji Nakagawa$^{1}$,\\
Yuichi Kasahara$^{5}$, Emil J. W. List-Kratochvil$^{2,6}$, Yusuke Kato$^{7}$, and Yoshihiro Iwasa$^{1,8\ast}$}


\maketitle 

\noindent
\textbf{1. Analogical explanation of vortex dynamics in BCS and BEC limits using charged particle motion}

For understanding of the vortex motion both in the BCS and BEC regimes, we compare the dynamics of a single vortex with that of a charged particle.
In the presence of a transport current $\bm{J}_\mathrm{tr}$, which is perpendicular to the direction of a magnetic field $\bm{B}$, a single vortex is subject to the driving force given as
\begin{equation}
    \bm{F}_\mathrm{d} = \bm{J}_\mathrm{tr} \times \bm{\phi}_0,
    \label{Eq:SVforce_d}
\end{equation}
where the vector $\bm{\phi}_0$ is parallel to $\bm{B}$ and has the modulus $|\bm{\phi}_0| = \phi_0=h/(2|e|)$ (flux quantum).
In addition to $\bm{F}_\mathrm{d}$, the vortex also feels the force due to environments such as interactions with impurities, imperfections, or phonons.
Assuming that the vortex moves with a velocity $\bm{v}_\mathrm{v}$, we can use two independent vectors $\bm{v}_\mathrm{v}$ and $\bm{v}_\mathrm{v} \times (\bm{\phi}_0 / \phi_0)$ to express the environmental force $\bm{F}_\mathrm{env}$:
\begin{equation}
    \bm{F}_\mathrm{env} = -\eta \bm{v}_\mathrm{v} + \eta' \bm{v}_\mathrm{v} \times (\bm{\phi}_0 / \phi_0),
    \label{Eq:SVforce_env}
\end{equation}
where $\eta$ ($> 0$) and $\eta'$ are transport coefficients for the vortex motion.
In Eq.~\eqref{Eq:SVforce_env}, $-\eta\bm{v}_\mathrm{v}$ represents a dissipative force since its work is negative ($-\eta\bm{v}_\mathrm{v} \cdot \bm{v}_\mathrm{v} < 0$), while $\eta'\bm{v}_\mathrm{v} \times (\bm{\phi}_0 / \phi_0)$ is non-dissipative since its work is always zero [$(\eta'\bm{v}_\mathrm{v} \times (\bm{\phi}_0 / \phi_0)) \cdot \bm{v}_\mathrm{v} = 0$].

The forces $\bm{F}_\mathrm{d}$ and $\bm{F}_\mathrm{env}$ are balanced as
\begin{equation}
    \bm{F}_\mathrm{d} + \bm{F}_\mathrm{env} = \bm{0}
    \label{Eq:forcebalance}
\end{equation}
in a steady flow of a single vortex.
In the BCS regime, since the dissipative force dominates the non-dissipative force due to the quasi-continuous spectrum in the vortex core (Fig.~1b in the main text), the force balance relation \eqref{Eq:forcebalance} leads to
\begin{equation}
    \bm{J}_\mathrm{tr} \times \bm{\phi}_0 - \eta\bm{v}_\mathrm{v} \approx \bm{0},
\end{equation}
which indicates that the vortex moves perpendicular to the transport current $\bm{J}_\mathrm{tr}$.
In the BEC regime, since the spectrum is quantized in the vortex and is gapful outside the core (Fig.~1a in the main text), there is no dissipation and the force balance relation \eqref{Eq:forcebalance} reduces to
\begin{equation}
    \bm{J}_\mathrm{tr} \times \bm{\phi}_0 + \eta'\bm{v}_\mathrm{v} \times (\bm{\phi}_0 / \phi_0) \approx \bm{0},
\end{equation}
which yields $\bm{v}_\mathrm{v} \approx - (\phi_0 / \eta') \bm{J}_\mathrm{tr}$, i.e. the vortex motion is anti-parallel to $\bm{J}_\mathrm{tr}$.

The force balance relation \eqref{Eq:forcebalance},
\begin{equation}
    \bm{F}_\mathrm{d} + \bm{F}_\mathrm{env} = \bm{J}_\mathrm{tr} \times \bm{\phi}_0 - \eta\bm{v}_\mathrm{v} + \eta'\bm{v}_\mathrm{v} \times (\bm{\phi}_0 / \phi_0) = \bm{0},
    \label{Eq:forcebalance2}
\end{equation}
for the vortex motion is less intuitive and thus it could be helpful to rewrite Eq.~\eqref{Eq:forcebalance2} in a dual picture so that an analogy with the motion of a charged particle is manifest.
Putting $\bm{\phi}_0 = q \phi_0 \bm{e}_z$ and $\bm{J}_\mathrm{tr} \times \bm{\phi}_0 =: q \bm{E}'$ with $q = \pm 1$, Eq.~\eqref{Eq:forcebalance2} becomes
\begin{equation}
    q \bm{E}' = \eta\bm{v}_\mathrm{v} - q \bm{v}_\mathrm{v} \times (\eta' \bm{e}_z).
    \label{Eq:forcebalance3}
\end{equation}
We see that Eq.~\eqref{Eq:forcebalance3} has the same form as the force balance relation of a charge $q$ in the presence of an ``electric field'' $\bm{E}'$ and a ``magnetic field'' $\bm{B}' = \eta' \bm{e}_z$.
When $\eta \ll |\eta'|$, which corresponds to the BEC regime, Eq.~\eqref{Eq:forcebalance3} describes the motion of a charged particle under a strong ``magnetic field'' $\bm{B}'$ and thus the motion is almost perpendicular to $q \bm{E}'$ and $\bm{v}_\mathrm{v}$ is anti-parallel to $\bm{J}_\mathrm{tr}$.
When $\eta \gg |\eta'|$, which corresponds to the BCS regime, Eq.~\eqref{Eq:forcebalance3} describes the motion of a charged particle under a weak ``magnetic field'' $\bm{B}'$ and thus the motion is almost parallel to $q \bm{E}'$ and $\bm{v}_\mathrm{v}$ is perpendicular to $\bm{J}_\mathrm{tr}$.
Retaining a small $|\eta'|$, we obtain a small Hall angle, the sign of which depends on that of $\eta'$, i.e. the direction of $\bm{B}'$.

\newpage

\noindent
\textbf{2. Unified experimental BCS-BEC crossover phase diagram}

The phase diagram of superconductors is usually drawn on the $T$-carrier density plane, whereas, in cold atom systems, the phase diagram is often drawn on the plane of $T$-$1/k_\mathrm{F} a_s$, where $k_\mathrm{F}$ and $a_s$ denote the Fermi vector and scattering length, respectively, and thus $1/k_\mathrm{F} a_s$ represents the normalized interaction strength. Therefore, it has not been possible to directly compare the two BCS-BEC crossover systems yet. Recently, we presented a phase diagram on the $T/T_\mathrm{F}$-$\Delta/E_\mathrm{F}$ plane for 2D superconductors \cite{Nakagawa2021}, which is free from the parameters specific to superconductors. On the other hand, in the 2D ${}^6 \mathrm{Li}$ system, the experimental determination of $\Delta/E_\mathrm{F}$ as a function of the interaction strength was recently reported \cite{Sobirey2021}. Combining the phase diagram on the same system published in 2015 \cite{Ries2015}, we are able to draw a phase diagram on the $T/T_\mathrm{F}$-$\Delta/E_\mathrm{F}$ plane for the 2D ${}^6 \mathrm{Li}$ system. This allows us to construct a unified experimental phase diagram of the BCS-BEC crossover, which is displayed in Supplementary Fig.~\ref{Fig:S01}. Though there remain discrepancies due to the difference in definition of each parameter, the phase diagram shows that the data of $\mathrm{Li}_x\mathrm{ZrNCl}$ and ${}^6 \mathrm{Li}$ just overlap with each other and encourages us to consider the BEC limit from the BCS side.
\\

\begin{figure}[h]
    \centering
    \includegraphics[scale=2.5]{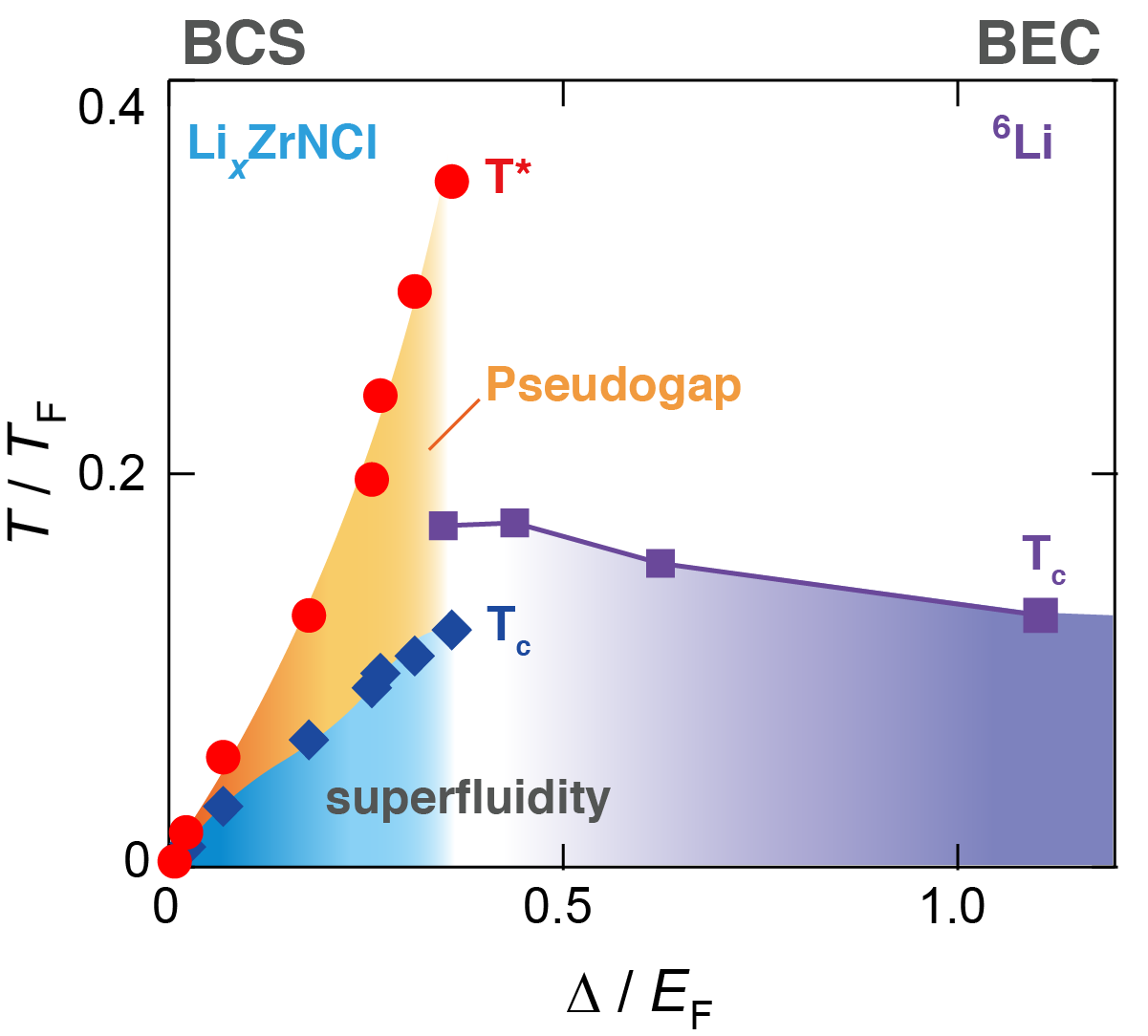}
    \caption{\textbf{A unified BCS-BEC crossover phase diagram from combined experimental data for cold atom superfluids $^{6}$Li and density-controlled superconductors Li$_{x}$ZrNCl.} Starting from the BCS side, the data points for the 2D superconductor Li$_{x}$ZrNCl \cite{Nakagawa2021}, the system studied in this work, are drawn. Here $T^{\ast}$ is the gap-opening temperature and $T_{\mathrm{c}}$ is the critical temperature. Starting from the BEC side, the purple data points correspond to the cold atom superfluid $^{6}$Li, taken from previous reports  \cite{Ries2015, Sobirey2021}. Drawing the phase diagram for both systems of superconductivity and cold atom superfluidity on a common scale reveals the achieved overlap of data points in the BCS-BEC crossover. This ultimately motivates the study of BEC superconductivty starting from the BCS regime in Li$_{x}$ZrNCl.}
    \label{Fig:S01}
\end{figure}

\noindent
\textbf{3. Basic transport properties and doping level determination}

To determine the doping level, i.e., Li content $x$, the Hall effect was used. The linear slope of Supplementary Fig.~\ref{Fig:S02}a was used to determine $x$ and a systematic dependency of the slope with changing Li ion concentration is apparent. The device operation for intercalation was analogous to previously established work \cite{Nakagawa2018}. To compute the Li content, the Hall coefficient at 150 K was measured and we assume that each Li ion supplies one electron to the ZrNCl system.
For the determined doping levels, superconductivity was achieved, and we show the longitudinal resistivity as a function of temperature for each in Supplementary Fig.~\ref{Fig:S02}b. With decreasing doping level, the critical temperature $T_\mathrm{c}$ increased from 11.4 K to 16.8 K for $x$ = 0.47 and 0.0040 respectively. $T_\mathrm{c}$ was determined by the temperature at which the resistivity is half of the normal-state value at 30 K. The superconducting transition is sharp for high doping levels but is significantly broadened towards lower doping levels. This may be explained by the enhanced fluctuation strength ($\beta$ in Supplementary Figs.~\ref{Fig:S06} and \ref{Fig:S07}) towards the BCS-BEC crossover. In addition, the dimensional crossover from an anisotropic 3D superconductor to a 2D superconductor can be relevant since the dimensional crossover occurs around $x\sim 0.1$, as discussed in previous works \cite{Nakagawa2018,Nakagawa2021}. In the low doping regime, the transition is better described by the Berezinskii-Kosterlitz-Thouless (BKT) transition using $T_\mathrm{BKT}$ instead of $T_\mathrm{c}$. However, the $T_\mathrm{c}$ values are sufficiently close to $T_\mathrm{BKT}$ to lend themselves for our further comparison.

\newpage

\begin{figure}[t]
    \centering
    \includegraphics[scale=1.2]{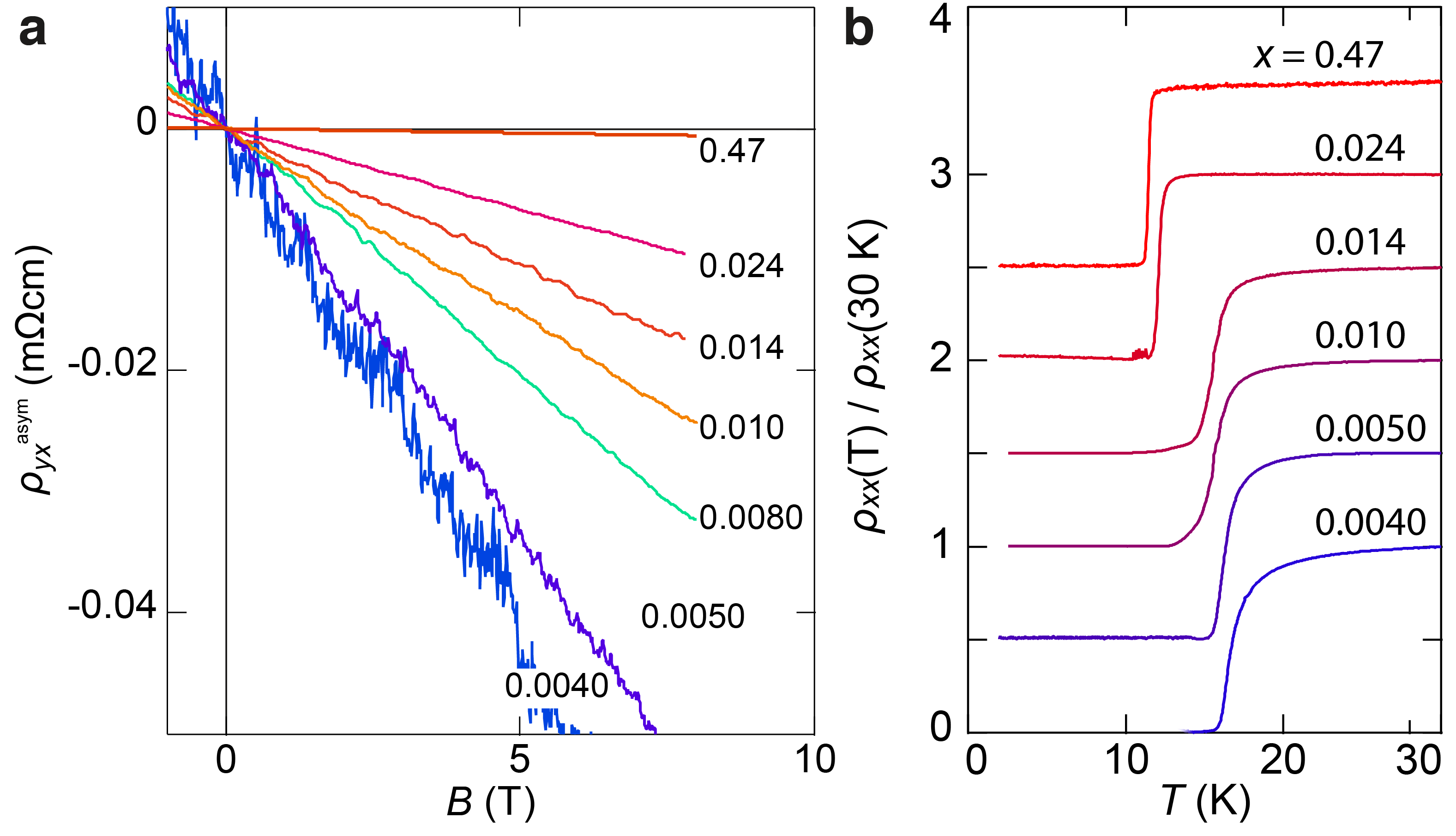}
    \caption{\textbf{Doping level determination and superconducting transitions.}\\
    (\textbf{a}) Anti-symmetrized transverse resistivity as a function of magnetic field measured at 150 K to determine the doping levels as annotated. The slope of the Hall signal is used to calculate $x$. (\textbf{b}) Doping dependence of the superconducting transition. Lower Li contents show higher critical temperatures $T_\mathrm{c}$ and broadened transitions due to superconducting fluctuation. Resistivity was normalized at 30 K and the curves are each shifted by 0.5. }
    \label{Fig:S02}
\end{figure}

\noindent
\textbf{4. Doping dependence of the upper critical field}

The out-of-plane upper critical field $B_\mathrm{c2}$ was determined by measuring the temperature dependence of resistivity at several magnetic fields. For the normal state, resistivity at high temperatures ($>30$ K) under the highest applied out-of-plane field of 8.8 T was chosen. The transition point is then defined as the half value of the normal state. These transition points are plotted as a function of temperature for each applied field at several doping levels in Supplementary Fig.~\ref{Fig:S03}a. Linear extrapolations to 0 K are plotted, which are used to determine $B_\mathrm{c2}$ at zero temperature. The enhancement of the upper critical field with decreased doping is evident, as also seen in Supplementary Fig.~\ref{Fig:S03}b, where the doping dependence of $B_\mathrm{c2}$ is shown. By using the Ginzburg-Landau (GL) model, 
$B_\mathrm{c2}(T) = (\phi_{0}/2\pi\xi^{2}) (1-T/T_\mathrm{c})$, where $\phi_{0}$ is the flux quantum, one can compute the   in-plane coherence length at zero temperature ($\xi$) by using the slope of the linear $B_\mathrm{c2}(T)$ relation. The values for each doping level are documented in Supplementary Table~\ref{Tab:Summary}. For decreasing doping, $\xi$ is decreasing. This indicates the realization of strongly coupled small Cooper pairs in the low-carrier density regime.
\\

\begin{figure}[h]
    \centering
    \includegraphics[scale=1.5]{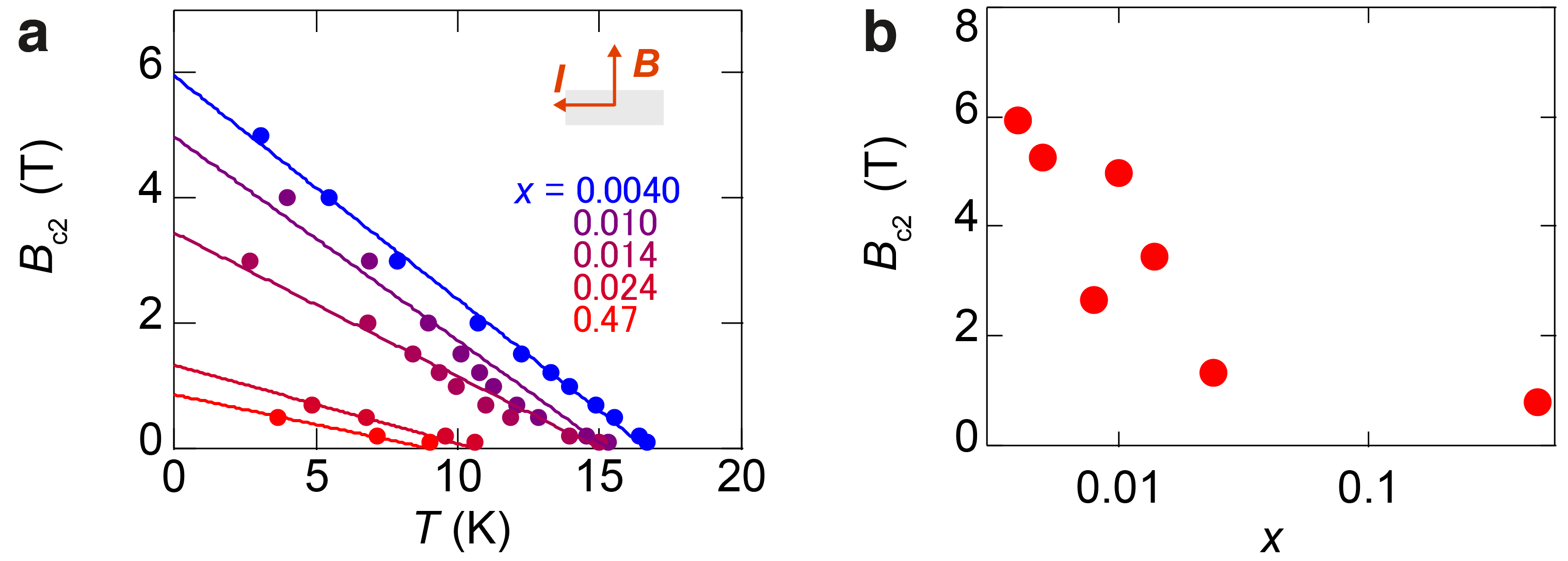}
    \caption{\textbf{Out-of-plane upper critical fields $B_{\mathrm{c2}}$.}\\
    (\textbf{a}) $B_\mathrm{c2}$ as a function of temperature at several doping levels. Solid lines represent linear extrapolations to 0 K for each doping level. (\textbf{b}) Doping dependence of $B_\mathrm{c2}$.}
    \label{Fig:S03}
\end{figure}

\noindent
\textbf{5. Summary of superconductivity properties at several doping levels}

In Supplementary Table~\ref{Tab:Summary}, the summarized values of the Li content $x$, carrier density $n$ at 150 K, Fermi energy $E_\mathrm{F}$, Fermi wave vector $k_\mathrm{F}$, critical temperature $T_\mathrm{c}$, out-of-plane upper critical field $B_\mathrm{c2}$  linearly extrapolated to 0 K, in-plane coherence length $\xi$ at 0 K, Hall mobility $\mu$ at 30 K and mean free path $\ell$ at 30 K for various achieved $x$ in Li$_{x}$ZrNCl are shown.
To calculate $E_\mathrm{F}$ and $k_\mathrm{F}$ from $n$, an ideal parabolic band dispersion in two-dimensions was employed. $k_\mathrm{F} = (4\pi n_\mathrm{layer}/ss')^{1/2}$ and $E_\mathrm{F} = \hbar^{2} k_\mathrm{F}^{2}/2m^{\ast}$, with $n_\mathrm{layer}$ the 2D carrier density per layer, $s$ the spin degree of freedom, $s'$ the valley degree of freedom, $\hbar$ the reduced Planck constant and $m^{\ast}$ the effective electron mass. In the case of Li$_{x}$ZrNCl, $s = s' = 2$ and $m^* = 0.9 m_{0}$, as reported before \cite{Kasahara2009}. Here $m_{0}$ denotes the free electron mass.
The Hall mobility was computed using $\mu = - \sigma_{xx} R_\mathrm{H}$, where $\sigma_{xx}$ is the longitudinal conductivity and $R_\mathrm{H}$ is the Hall coefficient. The mean free path was computed via $\ell = v_\mathrm{F} \mu\,  m^{\ast} q^{-1}$, where $q$ is the charge and $v_\mathrm{F}$ is the Fermi velocity computed via $v_\mathrm{F} = \hbar k_\mathrm{F}/m^{\ast}$.
\\

\begin{table}[h]
    \centering
    \caption{\textbf{Summary of parameters for doping-dependent superconductivity in Li$_{x}$ZrNCl.}}
    \vspace{0.2cm}
    \begin{tabular}{cccccccc}
        $x$& & $n(\mathrm{\times10^{20} \ cm^{-3}})$ & $E_\mathrm{F}$(meV) & $k_\mathrm{F}(\mathrm{nm^{-1}})$ & $T_\mathrm{c}$(K) & $B_\mathrm{c2}$(T) & $\xi$(nm) \\
        \hline
        $0.0040$ & & 0.718 & 8.8   & 0.456 & 16.8 & 5.95 & 7.44\\
        $0.0050$ & & 0.958 & 11.7 & 0.526 & 16.4 & 5.26 & 7.91\\
        $0.010$ & & 2.02  & 24.8 & 0.765 & 15.9 & 4.98 & 8.13\\
        $0.014$ & & 2.76  & 33.7 & 0.893 & 15.8 & 3.45 & 9.77\\
        $0.024$ & & 4.74  & 58.0 & 1.17 & 12.1 & 1.33 & 15.71\\
        $0.47$ & & 90.48  & 1107.0 & 5.11 & 11.4 & 0.79 & 20.47\\ \\
        $x$& & $\mu_\mathrm{30K}(\mathrm{cm^2 V^{-1} s^{-1}})$ & $\ell_\mathrm{30K}$(nm) \\
        \cline{1-4}
        $0.0040$ & & 68.42 & 5.79\\
        $0.0050$ & & 70.13 & 6.54\\
        $0.010$ & & 55.03 & 6.58\\
        $0.014$ & & 44.54 & 5.90\\
        $0.024$ & & 42.52 & 6.75\\
        $0.47$ & & 86.28 & 36.61\\
    \end{tabular}
    \label{Tab:Summary}
\end{table}

\begin{figure}[h]
    \centering
    \includegraphics[scale=2]{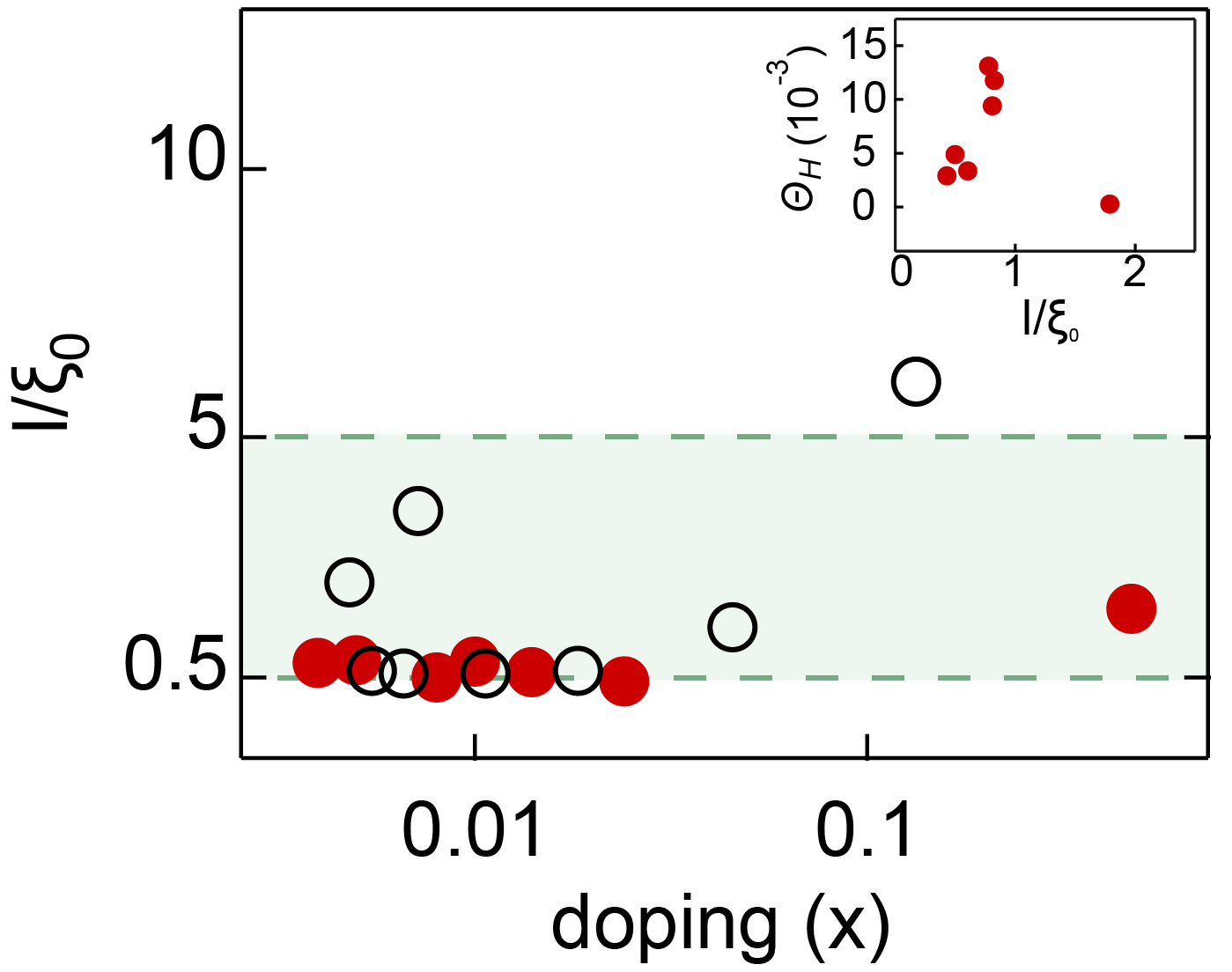}
    \caption{\textbf{Doping dependence of the ratio of the mean free path $l$ and superconducting coherence length $\xi_{0}$.}\\
   Red dots represent the data points from this work while black circles represent the data from the previous work~\cite{Nakagawa2021}. The green area highlights the moderately clean regime where $0.5 < l/\xi_{0} < 5$, above or below which the Hall anomaly is seemingly not observable as described by Hagen et al.~\cite{Hagen1993}. The inset shows the Hall angle dependence on $l/\xi_{0}$. Almost no modulation of the $l/\xi_{0}$ ratio with doping is observed, which is also reflected in the weak dependence of the Hall angle on this ratio. This concludes that the system stays in the relevant range of $l/\xi_{0}$ over the course of this work and the observed trend in the Hall angle vs doping is not dominated by a change of this ratio.}
    \label{Fig:S04}
\end{figure}

\noindent
\textbf{6. Theoretical determination of $T_\mathrm{c}$ by Hartree approximation}

As explained in Methods, the time-dependent Ginzuburg-Landau (TDGL) model is given by
\begin{equation}
    \left( \gamma + \mathrm{i} \lambda \right) \frac{\partial}{\partial t} \Delta (\bm{r}, t) = - \left[ \frac{T - T^*}{T^*} + b |\Delta (\bm{r}, t)|^2 - {\xi}^2 \left( \bm{\nabla} + \mathrm{i} \frac{2 \pi}{\phi_0} \bm{A} (\bm{r}) \right)^2 \right] \Delta (\bm{r}, t),
    \label{Eq:TDGL}
\end{equation}
where $\gamma = \pi / 8 T^*$, $\lambda = -(1 / 2 T^*) \partial T^* / \partial E_\mathrm{F}$, $\xi = \sqrt{\phi_0 / 2 \pi B_\mathrm{c2}(0)}$, $\bm{A} (\bm{r}) = B x \hat{y}$, and $\Delta (\bm{r}, t)$ is the superconducting order parameter varying in space and time.

We consider the GL Hamiltonian corresponding to Eq. (\ref{Eq:TDGL}) as
\begin{equation}
    H_\mathrm{GL} := a \int \mathrm{d}^2 \bm{r} \left[ \frac{T - T^*}{T^*} |\Delta (\bm{r})|^2 + \frac{b}{2} |\Delta (\bm{r})|^4 + {\xi}^2 \left| \left( \bm{\nabla} + \mathrm{i} \frac{2 \pi}{\phi_0} \bm{A} (\bm{r}) \right) \Delta (\bm{r}) \right|^2 \right],
    \label{Eq:GLHam}
\end{equation}
where $a$ is another phenomenological parameter.
Replacing $|\Delta|^4$ with $2 \langle |\Delta|^2 \rangle |\Delta|^2$ in Eq. (\ref{Eq:GLHam}) by the Hartree approximation, we obtain the approximated Hamiltonian
\begin{equation}
    H'_\mathrm{GL} := a \int \mathrm{d}^2 \bm{r} \left[ \epsilon |\Delta (\bm{r})|^2 + {\xi}^2 \left| \left( \bm{\nabla} + \mathrm{i} \frac{2 \pi}{\phi_0} \bm{A} (\bm{r}) \right) \Delta (\bm{r}) \right|^2 \right].
    \label{Eq:GLHam_Hartree}
\end{equation}
Here, the renormalized mass $\epsilon$ satisfies the self-consistent equation:
\begin{equation}
    \epsilon = \frac{T - T^*}{T^*} + b \langle |\Delta|^2 \rangle,
    \label{Eq:SCeq}
\end{equation}
where $\langle \cdots \rangle$ is the canonical average using the Hamiltonian $H'_\mathrm{GL}$ and the temperature $T$.
Expanding $\Delta (\bm{r})$ as $\Delta (\bm{r}) = \sum_{N, q} c_{N q} \varphi_{N q} (\bm{r})$ with the eigenfunction $\varphi_{Nq} (\bm{r}) \propto \mathrm{H}_N (x / l + q l) \exp [- (x / l + q l)^2 / 2 + \mathrm{i} q y]$, we can diagonalize $H'_\mathrm{GL}$ as $H'_\mathrm{GL} = a \sum_{N, q} [\epsilon + (2 N + 1) h] |c_{N q}|^2$.
Here, $N$ and $q$ are the Landau level index and its degeneracy index, respectively, $H_N (z)$ is the $N$th Hermite polynomial, $l := \sqrt{\phi_0 / 2 \pi B}$, and the dimensionless magnetic field is defined as $h := (\xi / l)^2 = B / B_\mathrm{c2}(0)$.
Then, the self-consistent equation (\ref{Eq:SCeq}) may be rewritten as
\begin{equation}
    \epsilon = \frac{T - T^*}{T^*} + \frac{b T h}{2 \pi a {\xi}^2} \sum_{N = 0}^{c / h} \frac{1}{\epsilon + (2 N + 1) h},
    \label{Eq:SCeq2}
\end{equation}
where $c$ is a cutoff parameter representing the limitation of the gradient expansion in the GL Hamiltonian (\ref{Eq:GLHam}).

We further rewrite Eq. (\ref{Eq:SCeq2}) as $\epsilon = (T - T^*) / T^* + (b T / 4 \pi a {\xi}^2) \{ [\sum_{N = 0}^{c / h} (N + \epsilon / 2h + 1 / 2)^{-1} - \ln (c / h)] + \ln (c / h) \}$.
Assuming large enough $c / h$ and taking $c / h \to \infty$ in $[\cdots]$ in the right-hand side, we can obtain
\begin{equation}
    \epsilon = \frac{T - T^*}{T^*} + \frac{b T}{4 \pi a {\xi}^2} \left[ - \psi \left( \frac{\epsilon + h}{2 h} \right) + \ln \frac{c}{h} \right],
    \label{Eq:SCeq3}
\end{equation}
where $\psi (z)$ is the digamma function, which satisfies $\psi (z) = \lim_{n \to \infty} [\ln n - \sum_{m=0}^n (m + z)^{-1}]$.
To express $c$ by the renormalized transition temperature, $T_\mathrm{c}$, we temporarily consider $h \to 0$ in Eq. (\ref{Eq:SCeq3}) and use $\psi(z) \to \ln z + O(z^{-1})$ for $z \to \infty$.
Then, we can obtain the equation for zero magnetic field as
\begin{equation}
    \epsilon = T \left[ \frac{1}{T^*} + \frac{b \ln (2 c)}{4 \pi a {\xi}^2} \right] - 1 - \frac{b T}{4 \pi a {\xi}^2} \ln \epsilon.
    \label{Eq:SCeq4}
\end{equation}
Based on the right-hand side of Eq. (\ref{Eq:SCeq4}), we define the renormalized $T_\mathrm{c}$ as 
\begin{equation}
    T_\mathrm{c} := \left[ \frac{1}{T^*} + \frac{b \ln (2 c)}{4 \pi a {\xi}^2} \right]^{-1}.
    \label{Eq:Tc}
\end{equation}
Note that the spontaneous symmetry breaking does not occur at finite temperatures since we consider a 2D system, and $T_\mathrm{c}$ characterizes a typical temperature for significant changes in physical quantities such as conductivity.

Using the expression of $T_\mathrm{c}$ [Eq. (\ref{Eq:Tc})] in Eq. (\ref{Eq:SCeq3}), we finally obtain the equation to determine $\epsilon$ as a function of $T / T_\mathrm{c}$ and $h \ [= B / B_\mathrm{c2}(0)]$:
\begin{equation}
    \epsilon = \frac{T - T_\mathrm{c}}{T_\mathrm{c}} + \beta \frac{T}{T_\mathrm{c}} \left[ - \psi \left( \frac{\epsilon + h}{2 h} \right) - \ln (2 h) \right].
    \label{Eq:SCeq5}
\end{equation}
Here, $\beta$ $(:= b T_\mathrm{c} / 4 \pi a {\xi}^2)$ is the dimensionless fluctuation interaction strength.
\\

\noindent
\textbf{7. Longitudinal and transverse conductivities}

In the following, we derive the electrical conductivity $\sigma_{xx}^\mathrm{V}$ and $\sigma_{yx}^\mathrm{V}$ due to the vortex liquid using the Hartree approximation \cite{Ullah1991}.
According to the linear response theory \cite{Abrahams1971}, we can express $\sigma_{ab}^\mathrm{V}$ as
\begin{equation}
    \sigma_{a b}^\mathrm{V} = \frac{1}{TS} \int_0^\infty \mathrm{d} t \int \mathrm{d}^2 \bm{r} \int \mathrm{d}^2 \bm{r}' \langle j_a (\bm{r}, t) j_b (\bm{r}', 0) \rangle,
    \label{Eq:LinearResponse}
\end{equation}
where $S$ is the total area of the system, and $\bm{j} (\bm{r}, t)$ is the local current density defined as
\begin{equation}
    \left. \bm{j} (\bm{r}, t) := \frac{2 \pi}{\phi_0} a {\xi}^2 \left\{ \mathrm{i} (\bm{\nabla}_1 - \bm{\nabla}_2)  - \frac{2\pi}{\phi_0} [\bm{A} (\bm{r}_1) + \bm{A} (\bm{r}_2)] \right\} \Delta (\bm{r}_1, t) \Delta^* (\bm{r}_2, t) \right|_{\bm{r}_1 = \bm{r}_2 = \bm{r}}.\label{eq: Js-def}
\end{equation}

Within the Hartree approximation \cite{Ullah1991}, $\langle \cdots \rangle$ in Eq. (\ref{Eq:LinearResponse}) means the canonical average using $H'_\mathrm{GL}$ [Eq. (\ref{Eq:GLHam_Hartree})] for the initial state.
Using $\epsilon$, which satisfies Eq. (\ref{Eq:SCeq5}), the TDGL model (\ref{Eq:TDGL}) is approximated as
\begin{equation}
    \left( \gamma + \mathrm{i} \lambda \right) \frac{\partial}{\partial t} \Delta (\bm{r}, t) = - \left[ \epsilon - {\xi}^2 \left( \bm{\nabla} + \mathrm{i} \frac{2 \pi}{\phi_0} \bm{A} (\bm{r}) \right)^2 \right] \Delta (\bm{r}, t).
    \label{Eq:TDGL_Hartree}
\end{equation}
Expanding $\Delta (\bm{r}, t)$ as $\Delta (\bm{r}, t) = \sum_{N, q} c_{N q} (t) \varphi_{N q} (\bm{r})$, we can solve Eq. (\ref{Eq:TDGL_Hartree}) as $c_{N q} (t) = c_{N q} \exp [- (\epsilon + h + 2 h N) t / (\gamma + \mathrm{i} \lambda)]$ with $c_{N q} := c_{N q} (0)$.
Thus, Eq. (\ref{Eq:LinearResponse}) leads to
\begin{eqnarray}
    \sigma_{a b}^\mathrm{V} = \frac{1}{TS} \left( \frac{2 \pi}{\phi_0} a {\xi}^2  \right)^2 \sum_{N, N', q, q'} \langle |c_{Nq}|^2 \rangle \langle |c_{N' q'}|^2 \rangle \int_0^\infty \mathrm{d} t \, \mathrm{e}^{-\Gamma (N, N') t} \nonumber \\
    \times \int \mathrm{d}^2 \bm{r} \int \mathrm{d}^2 \bm{r}' [\hat{\bm{\Pi}}_1 + \hat{\bm{\Pi}}_2^*]_a [\hat{\bm{\Pi}}'_1 + \hat{\bm{\Pi}}^{\prime *}_2]_b \varphi_{N q} (\bm{r}_1) \varphi^*_{N' q'} (\bm{r}_2) \varphi_{N' q'} (\bm{r}'_1) \varphi^*_{N q} (\bm{r}'_2) |_{\bm{r}^{(\prime)}_{1} = \bm{r}^{(\prime)}_{2} = \bm{r}^{(\prime)}},
    \label{Eq:LinearResponse2}
\end{eqnarray}
where $\Gamma (N, N') := 2 \{ \gamma [\epsilon + (N_1 + N_2 + 1) h] - \mathrm{i} \lambda (N_1 - N_2) h \} / (\gamma^2 + \lambda^2)$, and $\hat{\bm{\Pi}}_i^{(\prime)}$ [$:= -\mathrm{i} \bm{\nabla}_i^{(\prime)} + (2 \pi / \phi_0) \bm{A} (\bm{r}_i^{(\prime)})$] is the gauge-invariant momentum operator.
Using the ladder operators $\hat{a}_q := (x + q l^2 + l^2 \partial_x) / \sqrt{2} l$ and $\hat{a}_q^\dag := (x + q l^2 - l^2 \partial_x) / \sqrt{2} l$, we can obtain the relations such as $\hat{\Pi}_x \varphi_{N q} (\bm{r}) = - (\mathrm{i} / \sqrt{2} l) (\hat{a}_q - \hat{a}_q^\dag) \varphi_{N q} (\bm{r}) = - (\mathrm{i} / \sqrt{2} l) [\sqrt{N} \varphi_{N-1, q} (\bm{r}) - \sqrt{N + 1} \varphi_{N + 1, q} (\bm{r})]$.
Performing the time and space integrations and the canonical average in Eq. (\ref{Eq:LinearResponse2}), we can finally obtain the following formulas:
\begin{eqnarray}
    && \sigma_{xx}^\mathrm{V} = \sigma_{yy}^\mathrm{V} = \frac{\gamma T h^2}{\pi} \left( \frac{2 \pi}{\phi_0} \right)^2 \left( 1 + \frac{\lambda^2}{\gamma^2} \right) \sum_{N = 0}^{c/h} \frac{(N+1) \mu_{N + 1/2}}{\mu_N \mu_{N+1} ({\mu_{N+1/2}}^2 + \lambda^2 h^2 / \gamma^2)},
    \label{Eq:sigmaxx} \\
    && \sigma_{yx}^\mathrm{V} = - \sigma_{xy}^\mathrm{V} = - \frac{\lambda T h^3}{\pi} \left( \frac{2 \pi}{\phi_0} \right)^2 \left( 1 + \frac{\lambda^2}{\gamma^2} \right) \sum_{N = 0}^{c/h} \frac{N+1}{\mu_N \mu_{N+1} ({\mu_{N+1/2}}^2 + \lambda^2 h^2 / \gamma^2)},
    \label{Eq:sigmayx}
\end{eqnarray}
where $\mu_N := \epsilon + h + 2 N h$, and a cutoff $c$ is introduced similarly to Eq. (\ref{Eq:SCeq2}).

For $\epsilon + h \ll h$, the $N=0$ terms are dominant in Eqs. (\ref{Eq:sigmaxx}) and (\ref{Eq:sigmayx}), and we obtain the asymptotic expressions as
\begin{equation}
    \sigma_{xx}^\mathrm{V} = \sigma_{yy}^\mathrm{V} \simeq \frac{2 \pi \gamma T}{{\phi_0}^2 (\epsilon + h)} \ \ \ \ \ (\mathrm{for} \ \epsilon + h \ll h),
    \label{Eq:sigmaxxlowT}
\end{equation}
\begin{equation}
    \sigma_{yx}^\mathrm{V} = - \sigma_{xy}^\mathrm{V} \simeq -\frac{2 \pi \lambda T}{{\phi_0}^2 (\epsilon + h)} \ \ \ \ \ (\mathrm{for} \ \epsilon + h \ll h).
    \label{Eq:sigmayxlowT}
\end{equation}
Thus, for low enough temperatures, where $\epsilon + h \ll h$ and $\sigma_{ab}^\mathrm{V}$ dominates over the normal-state conductivity $\sigma_{ab}^\mathrm{N}$, the resistivities are given as $\rho_{xx} := (\sigma_{xx}^\mathrm{N} + \sigma_{xx}^\mathrm{V}) / [(\sigma_{xx}^\mathrm{N} + \sigma_{xx}^\mathrm{V})^2 + (\sigma_{yx}^\mathrm{N} + \sigma_{yx}^\mathrm{V})^2] \simeq \sigma_{xx}^\mathrm{V} / [(\sigma_{xx}^\mathrm{V})^2 + (\sigma_{yx}^\mathrm{V})^2]$ and $\rho_{yx} := -(\sigma_{yx}^\mathrm{N} + \sigma_{yx}^\mathrm{V}) / [(\sigma_{xx}^\mathrm{N} + \sigma_{xx}^\mathrm{V})^2 + (\sigma_{yx}^\mathrm{N} + \sigma_{yx}^\mathrm{V})^2] \simeq -\sigma_{yx}^\mathrm{V} / [(\sigma_{xx}^\mathrm{V})^2 + (\sigma_{yx}^\mathrm{V})^2]$, and the Hall angle $\mathit{\Theta}_\mathrm{H}$ follows
\begin{equation}
    \tan \mathit{\Theta}_\mathrm{H} := \rho_{yx} / \rho_{xx} \simeq -\sigma_{yx}^\mathrm{V} / \sigma_{xx}^\mathrm{V} \simeq \lambda / \gamma. \ \ \ \ \ (\mathrm{for} \ \epsilon + h \ll h, \ \sigma_{xx}^\mathrm{N} \ll \sigma_{xx}^\mathrm{V}, \ \mathrm{and} \ |\sigma_{yx}^\mathrm{N}| \ll |\sigma_{yx}^\mathrm{V}|)
    \label{Eq:HallAngle}
\end{equation}
By explicitly setting $\gamma = \pi / 8T^*$ and $\lambda = -(\partial T^* / \partial E_\mathrm{F}) / 2 T^*$ in (\ref{Eq:HallAngle}), we finally obtain
\begin{equation}
    \tan \mathit{\Theta}_\mathrm{H} \simeq -(4 / \pi) \partial T^* / \partial E_\mathrm{F}.
    \label{Eq:HallAngle2}
\end{equation}
\\

\noindent
\textbf{8. TDGL equation and Hall conductivity}

We discuss the crucial role of $\lambda$ in the left-hand-side of Eq.~\eqref{Eq:TDGL}. When $\lambda=0$, Eq.~\eqref{Eq:TDGL} has a particle-hole symmetry, i.e, When $\Delta$ and $\bm{A}$ satisfy the Eq.~\eqref{Eq:TDGL} with $\lambda=0$, $\Delta'=\Delta^*$ and $\bm{A}'=-\bm{A}$ do the same equation. Both the current density and electric field change their sign and it thus follows that $\sigma_{xy}(\bm{B})=\sigma_{xy}(-\bm{B})$ under this transformation for $\bm{B}$ parallel to $z$-axis. The Onsager relation $\sigma_{xy}(\bm{B})=\sigma_{yx}(-\bm{B})$ together with the rotational symmetry $\sigma_{xy}(\bm{B})=-\sigma_{yx}(\bm{B})$ in the $xy$ plane yields $\sigma_{xy}(\bm{B})=-\sigma_{xy}(-\bm{B})$. We thus see that $\sigma_{xy}(\bm{B})=0$ when $\lambda=0$. 

As we will confirm in the following calculation, the sign of the $\lambda$ determines that of the Hall conductivity. Before explicit calculation, we discuss the sign of the Hall conductivity in an intuitive way. For simplicity, we set $\gamma=0$, which is irrelevant to the sign of the Hall conductivity, then Eq.~\eqref{Eq:TDGL} reduces to the form of non-linear Schr\" odinger equation, where $\lambda$ corresponds to $-m^*\xi^2/\hbar$ with the inertial mass $m^*$ of a Cooper pair. When $\lambda<0$, the Eq.~\eqref{Eq:TDGL} with $\gamma=0$ describes the dynamics of charged condensate with a positive inertial mass and the resultant Hall conductivity with the same sign as the electrons in the normal state. When $\lambda>0$, on the other hand, the same equation describes the dynamics of charged condensate with a negative  mass. In this case, the Hall effect due to motion of condensate has the opposite sign to that in the normal state. 

To gain further insight into the dynamics of the condensate, it would be helpful to   
to rewrite $\Delta(\bm{r},t)$ as $|\Delta(\bm{r},t)|e^{i \chi(\bm{r},t)}$ and decompose TDGL equation \eqref{Eq:TDGL} multiplied by $e^{-i \chi(\bm{r},t)}$ into real and imaginary parts
\begin{subequations}
\begin{align}
\gamma \frac{\partial |\Delta|}{\partial t} &=-\frac{1}{2a}\frac{\delta H_{\rm GL}}{\delta |\Delta|}+\lambda |\Delta|\frac{\partial \chi}{\partial t}\label{eq: TDGL-re}\\
\frac{\partial \rho_{\rm s}}{\partial t}+\nabla\cdot \bm{j}_{\rm s}&
=-\frac{4\pi\gamma|\Delta|^2}{\phi_0}\frac{\partial \chi}{\partial t}.\label{eq: TDGL-im}
\end{align}
\end{subequations}
The GL Hamiltonian \eqref{Eq:GLHam} is introduced in eq.~\eqref{eq: TDGL-re}, which describes the relaxation dynamics of the condensate. In eq.~\eqref{eq: TDGL-im}, we introduce the notation: 
\begin{equation}
    \rho_{\rm s}:=\frac{2\pi a\lambda|\Delta|^2}{\phi_0},\quad
    \bm{j}_{\rm s}=-2\left(\frac{2\pi|\Delta|\xi}{\phi_0}\right)^2\underbrace{\left(\bm{A}+\frac{\phi_0}{2\pi}\nabla\chi\right)}_{=:\bm{Q}},\label{eq: rho-s-j-s}
\end{equation}
the latter of which is nothing but \eqref{eq: Js-def}. We can regard Eq.\eqref{eq: TDGL-im} as the equation of continuity of the superfluid component of charge and current with the sink/source term. Conserved is the sum of the superfluid component and normal component of charge density. We thus interpret the righ-hand side of eq.~\eqref{eq: TDGL-im} as the conversion rate of charge density from the normal to superfluid component and introduce the notation 
\begin{equation}
\left[\frac{d\rho_{\rm s}}{d t}\right]_{\rm conv}:=-\frac{4\pi\gamma|\Delta|^2}{\phi_0}\frac{\partial \chi}{\partial t}.\label{eq: conversion-rate}
\end{equation}
In Eq.~\eqref{eq: rho-s-j-s} for $\lambda<0$, $\rho_{\rm s}$ is negative and corresponds to the positive electron density. Thus the dynamics of the condensate is similar to the electron motion. 
In Eq.~\eqref{eq: rho-s-j-s} for $\lambda>0$,  $\rho_{\rm s}$ is positive and corresponds to the deficit of electron number density. We then expect the dynamics of the condensate is similar to that of holes. 
Further we can discuss the dynamics of the condensate on the basis of momentum balance relation, which corresponds to the Euler equation (equation of motion ) in hydrodynamics
\begin{equation}
\frac{\partial(-\rho_{\rm s}\bm{Q})}{\partial t}+\bm{\nabla}\cdot\bm{\mathcal{P}}
=
\rho_{\rm s}\bm{\varepsilon}+\bm{j}_{\rm s}\times\bm{h}+2a\gamma \frac{\partial|\Delta|}{\partial t}\bm{\nabla}|\Delta|-\left[\frac{d\rho_{\rm s}}{d t}\right]_{\rm conv}\bm{Q},\label{eq: momentum-balance}
\end{equation}
which follows from Eqs.~\eqref{eq: TDGL-re} and \eqref{eq: TDGL-im}, and the Ampere-Maxwell equation (see derivation of \cite{Kato2016}). Let us see the physical meaning of each term in order to confirm that this equation is really regarded as the momentum balance relation. We start with the right-hand side. Here the electric magnetic fields are denoted by $\bm{\varepsilon}=-\partial \bm{A}/\partial t$ and 
$\bm{h}=\bm{\nabla}\times\bm{A}$. The first two terms in the right-hand side represent the electromagnetic Lorentz force. The third term in the right-hand side is the dissipation force due to the time variation of the modulus of $|\Delta|$ (This mechanism was first pointed out by Tinkham~\cite{Tinkham1964}). The last term in the right-hand side in Eq. \eqref{eq: momentum-balance} is the other dissipation force due to conversion between the superfluid and normal components. This disspative force is caused by the time-variation of the phase of $\Delta$. Thus these two terms show that the vortex motion is the source of the dissipative force. 
In the left-hand side, $\bm{\mathcal{P}}$ represents the hydrodynamic momentum flux tensor, which is given in the present case by 
\begin{align}
(\bm{\mathcal{P}})_{\mu\nu}&=-j_{{\rm s},\mu}Q_\nu+2a \xi^2 \partial_\mu |\Delta|\partial_\nu |\Delta|-\delta_{\mu\nu}\left(a\mathcal{F}-\rho_{\rm s}\frac{\phi_0}{2\pi}\frac{\partial \chi}{\partial t}\right)\\
\mathcal{F}&=
\frac{T - T^*}{T^*}|\Delta|^2 + \frac{b}{2} |\Delta |^4 +{\xi}^2 \left( \bm{\nabla}|\Delta|\right)^2 +\left(\frac{2 \pi\xi}{\phi_0}|\Delta|\bm{Q}  \right)^2.
\end{align}
These expressions in the London limit, where  $|\Delta|$ is spatially uniform, reduce to 
\begin{align}
(\bm{\mathcal{P}})_{\mu\nu}&\rightarrow
\frac12\left(\frac{\phi_0}{2\pi |\Delta|\xi}\right)^2\left(j_{{\rm s},\mu}j_{{\rm s},\nu}-\frac{\delta_{\mu\nu}\bm{j}_{\rm s}^2}{2}\right),
\end{align}
which coincides with the momentum flux tensor in the London equation~\cite{London1950}. We then finally identify $-\rho_{\rm s}\bm{Q}$ in the first term in the left-hand side with the superfluid component of the momentum density. We see that this terms has the same sign as that of $\rho_{\rm s}\bm{j}_{\rm s}$. When $\lambda<0$, the momentum density has antiparallel to the electric current density $\bm{j}_{\rm s}$ and thus the condensate corresponds to the positive electron density and the dynamics is similar to the electron motion. When $\lambda>0$, on the other hand, the momentum density has parallel to the electric current density $\bm{j}_{\rm s}$ and thus the condensate corresponds to deficit of electron density and the dynamics is similar to the hole motion.   
In this section, we argue that the dynamics of the condensate described by the TDGL equation is similar to that of electron (hole) when $\lambda$ is negative (positive) on the basis of analogy with Schr\"odinger equation, Eq.~\eqref{eq: TDGL-im}, and Eq.~\eqref{eq: momentum-balance}.

Our expectation on the relation between the sign of the Hall conductivity and that of $\lambda$ is consistent with the results of the Hartree approximation. 
\\

\noindent
\textbf{9. Comparison of theory and experiment}

\begin{figure}[h]
    \centering
    \includegraphics[scale=0.7]{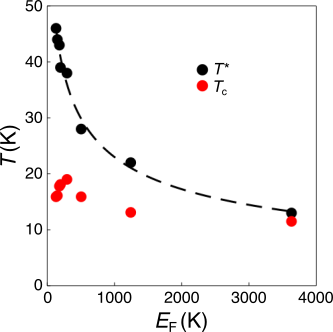}
    \caption{\textbf{Gap-opening temperature $T^*$ and superconducting critical temperature $T_\mathrm{c}$ as previously established\cite{Nakagawa2021}.}\\
    The black dashed line is the fitting curve of $T^* (E_\mathrm{F})$.}
    \label{Fig:S05}
\end{figure}

We examined whether the theoretical expressions of the conductivities [Eqs. (\ref{Eq:sigmaxx}) and (\ref{Eq:sigmayx})] and the Hall angle [Eq. (\ref{Eq:HallAngle2})] can explain the experimentally observed temperature and field dependence of $\rho_{xx}$ and $\rho_{yx}$ (Figs. 2a and 2c) and the concentration dependence of the Hall angle (Fig. 3b).
The phenomenological parameters [$T^*$, $\partial T^* / \partial E_\mathrm{F}$, $T_\mathrm{c}$, $B_\mathrm{c2} (0)$, $\sigma_{xx}^\mathrm{N}$, $\sigma_{yx}^\mathrm{N} / B$] were obtained from the present and previous \cite{Nakagawa2021} experiments (Supplementary Table~\ref{Tab:ParamsTheory}), and the dimensionless fluctuation interaction $\beta$ was set by hand.
We fitted the previous data of $T^* (E_\mathrm{F})$ \cite{Nakagawa2021} with a function $f(z) := c_1 / (1 + c_2 z^{c_3})$ (Supplementary Fig.~\ref{Fig:S05}), where the best-fitted parameters are $(c_1, c_2, c_3) = (97.8, 0.0888, 0.521)$, and we extrapolated the fitting curve to obtain $T^*$ and $\partial T^* / \partial E_\mathrm{F}$ for the $E_\mathrm{F}$ values corresponding to $x = 0.0040$ and $0.47$ in the present experiment.
Note that $T^*$ is estimated smaller than $T_\mathrm{c}$ for $x = 0.47$ simply due to the curve fitting to a few data points (Supplementary Fig.~\ref{Fig:S05}), which will not qualitatively affect the outcomes, though $\sigma_{xx}^\mathrm{V}$ and $\sigma_{yx}^\mathrm{V}$ may be overestimated.
$T_\mathrm{c}$ and $B_\mathrm{c2}$ were obtained from the resistance measurement as the point for half of the normal resistance.
$\sigma_{xx}^\mathrm{N}$ and $\sigma_{yx}^\mathrm{N} / B$ were determined from the experimental data at 30 K.
In Fig. 4c, we plotted $\mathit{\Theta}_\mathrm{H}$ for low enough temperatures, based on the fitting curve of $T^* (E_\mathrm{F})$ (Supplementary Fig.~\ref{Fig:S05}) and $\tan \mathit{\Theta}_\mathrm{H} \simeq -(4 / \pi) \partial T^* / \partial E_\mathrm{F}$ [Eq. (\ref{Eq:HallAngle2})].

Using the parameters in Supplementary Table~\ref{Tab:ParamsTheory}, we calculated $\rho_{xx}$ and $\rho_{yx}$ based on Eqs. (\ref{Eq:sigmaxx}) and (\ref{Eq:sigmayx}), without cutoff ($c / h \to \infty$) for simplicity.
The obtained temperature and field dependence of resistivity is shown in Supplementary Figs.~\ref{Fig:S06} ($x = 0.004$) and \ref{Fig:S07} ($x = 0.47$) for $\beta = 10^{-4}$, $10^{-3}$, and $10^{-2}$.
For $x = 0.004$ (Supplementary Fig.~\ref{Fig:S06}), the experimentally observed sign reversal and positive peak of $\rho_{yx}$ (Fig. 2a) are qualitatively reproduced by the theory, in a broad range of the fluctuation strength parameter $\beta$.
The peaks of $\rho_{yx}$ are quantitatively higher than the experimentally observed ones especially at high fields, which might be explained by theoretical overestimation of the fluctuation contribution at high fields as known for the fluctuation-induced diamagnetism \cite{Lee1971,Lee1972} (see also \cite{Carballeira2000,Vidal2002} for the overestimation of the fluctuation effects without introducing a cutoff in the GL formalism).
On the other hand, for  $x = 0.47$ (Supplementary Fig.~\ref{Fig:S07}), $\rho_{yx}$ does not show a clear peak regardless of $\beta$, consistently with the experimental results (Fig. 2c).
More quantitative comparison will require calculations starting from a microscopic Hamiltonian (e.g., \cite{Shi2021}).
For Figs. 4a and 4b, we chose $\beta = 10^{-3}$ (Supplementary Fig.~\ref{Fig:S06}, center) and $\beta = 10^{-4}$ (Supplementary Fig.~\ref{Fig:S07}, left), respectively.
\\

\begin{table}[h]
    \centering
    \caption{\textbf{Parameters used to produce Figs. 4a--c in the main text.}}
    \vspace{0.2cm}
    \begin{tabular}{cccccc}
        $x$ & & $E_\mathrm{F}$ (K) & $T^*$(K) & $\partial T^* / \partial E_\mathrm{F}$ & $T_\mathrm{c}$(K)\\
        \hline
        $0.0040$ & & 102 & 49.1 & $-0.125$ & 16.8\\
        $0.47$ & & $1.28 \times 10^4$ & 7.35 & $-2.76 \times 10^{-4}$ & 11.4\\ \\
        $x$ & & $B_\mathrm{c2} (0)$(T) & $\sigma_{xx}^\mathrm{N}$($\Omega^{-1} \mathrm{cm}^{-1}$) & $\sigma_{yx}^\mathrm{N} / B$ ($\Omega^{-1} \mathrm{cm}^{-1} T^{-1}$) &\\
        \cline{1-5}
        $0.0040$ & & 5.95 & 670 & 4.52 &\\
        $0.47$ & & 0.79 & $6.57 \times 10^4$ & $547$ &
    \end{tabular}
    \label{Tab:ParamsTheory}
\end{table}

\begin{figure}[h]
    \centering
    \includegraphics[scale=0.6]{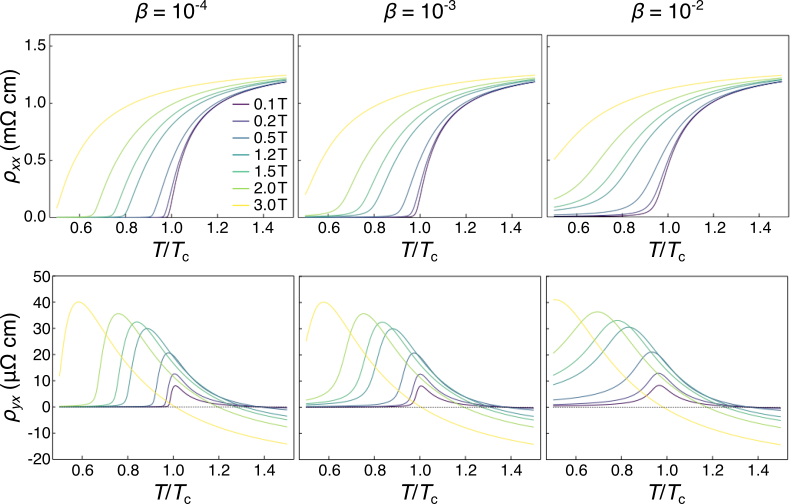}
    \caption{\textbf{Theoretical longitudinal and transverse resistivities for $x = 0.0040$.}\\
    The left, center, and right panels correspond to $\beta = 10^{-4}$, $10^{-3}$, and $10^{-2}$, respectively.
    Other parameters are summarized in Table S2.}
    \label{Fig:S06}
\end{figure}

\clearpage

\begin{figure}[h]
    \centering
    \includegraphics[scale=0.6]{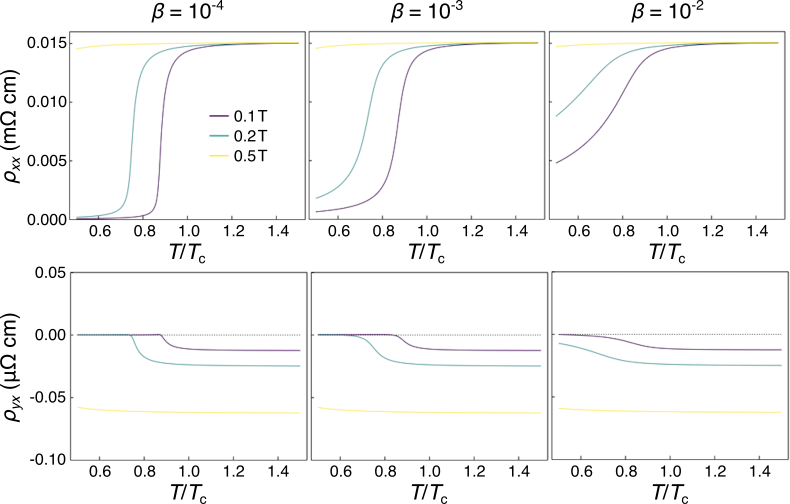}
    \caption{\textbf{Theoretical longitudinal and transverse resistivities for $x = 0.47$.}\\
    The left, center, and right panels correspond to $\beta = 10^{-4}$, $10^{-3}$, and $10^{-2}$, respectively.
    Other parameters are summarized in Table S2.}
    \label{Fig:S07}
\end{figure}

\noindent
\textbf{10. Dilute Fermi gas model}

If we further reduce the doping $x$ of Li$_x$ZrNCl than in the present experiment, the distance between conduction electrons can become larger than the range of the effective attractive interaction.
Then, the system may be described by the 2D Fermi gas model with a contact attractive interaction ($g > 0$):
\begin{equation}
    H := \sum_{\sigma = \uparrow, \downarrow} \int \mathrm{d}^2 \bm{r} \, \psi^\dag_\sigma (\bm{r}) \left( - \frac{\bm{\nabla}^2}{2 m} \right) \psi_\sigma (\bm{r}) - g \int \mathrm{d}^2 \bm{r} \, \psi^\dag_\uparrow (\bm{r}) \psi^\dag_\downarrow (\bm{r}) \psi_\downarrow (\bm{r}) \psi_\uparrow (\bm{r}),
    \label{Eq:Hamiltonian}
\end{equation}
where $\psi_\sigma (\bm{r})$ and $\psi_\sigma^\dag (\bm{r})$ are the Fermion field operators.
In this model, the crossover from the BCS regime to the BEC regime occurs as the Fermion density is decreased \cite{Eagles1969,Gusynin1999,Botelho2006}.
Based on the expansion of the fluctuation propagator \cite{Melo1993,Yanase1999,Stajic2003,Han2011}, we can derive the TDGL model corresponding to Eq. (\ref{Eq:Hamiltonian}) in the form of Eq. (\ref{Eq:TDGL}).
The coefficients of the TDGL model are now connected to the microscopic quantities as
\begin{equation}
    \gamma = \frac{\pi}{4 T^* c} \theta (\mu^*),
    \label{Eq:TDGL_gamma}
\end{equation}
\begin{equation}
    \lambda = - \frac{1}{4 T^* c} \mathrm{P} \int_{\mu^* / 2 T^*}^\infty \mathrm{d} x \frac{\tanh x}{x^2},
    \label{Eq:TDGL_lambda}
\end{equation}
\begin{equation}
    b = \frac{m}{8 \pi c} \int_0^\infty \mathrm{d} \varepsilon \left[ \frac{X}{(\varepsilon - \mu^*)^3} - \frac{Y}{2 T^* (\varepsilon - \mu^*)^2} \right],
\end{equation}
\begin{equation}
    {\xi}^2 = -\frac{1}{32 \pi c} \int_0^\infty \mathrm{d} \varepsilon \left[ \frac{X}{(\varepsilon - \mu^*)^2} - \frac{Y}{2 T^* (\varepsilon - \mu^*)} + \frac{\varepsilon X Y}{2 (T^*)^2 (\varepsilon - \mu^*)} \right],
\end{equation}
where $\theta (z)$ is the Heaviside step function, $\mathrm{P} (\cdots)$ means the Cauchy principal value, $\mu^*$ is the mean-field chemical potential at $T^*$, $X := \tanh [(\varepsilon - \mu^*) / 2 T^*]$, $Y := 1 - X^2$, $c := 1 + (E_\mathrm{F} / \mu^*) \exp [(E_\mathrm{F} - \mu^*) / T^*] \tanh (\mu^*/2 T^*)$, and $E_\mathrm{F}$ is the Fermi energy.

In the BCS side (high density and $\mu^* > 0$), Eqs. (\ref{Eq:TDGL_gamma}) and (\ref{Eq:TDGL_lambda}) lead to $\gamma > 0$ and $\lambda < 0$, and thus $\mathit{\Theta}_\mathrm{H} < 0$ from Eq. (\ref{Eq:HallAngle}).
Especially, deep in the BCS regime, where $T^* \ll \mu^* \simeq E_\mathrm{F}$, we can see $c \simeq 2$, $\gamma \simeq \pi / 8 T^*$, $\lambda \simeq - 1 / 4 E_\mathrm{F}$, and thus $\mathit{\Theta}_\mathrm{H} \simeq -2T^* / \pi E_\mathrm{F}$.
Note that the sign of $\mathit{\Theta}_\mathrm{H}$ (negative) is opposite to the experimentally observed values (positive), which suggests that the contact interaction in Eq. (\ref{Eq:Hamiltonian}) cannot explain the present doping range, and finite-range interactions may be important as indicated by a recent work \cite{Shi2021}.

In the BEC side (low density and $\mu^* < 0$), we see $\gamma = 0$ from Eq. (\ref{Eq:TDGL_gamma}), and thus $\sigma_{xx}^\mathrm{V} = 0$ according to Eq. (\ref{Eq:sigmaxx}).
For low temperatures satisfying $\epsilon + h \ll h$, the normal-state conductivity, $\sigma_{ab}^\mathrm{N}$, is expected to be negligible compared to $\sigma_{yx}^\mathrm{V} \simeq - 2 \pi \lambda T / [{\phi_0}^2 (\epsilon + h)]$ [Eq. (\ref{Eq:sigmayxlowT})].
Thus, the Hall angle will be $\mathit{\Theta}_\mathrm{H} = -\arctan [(\sigma_{yx}^\mathrm{N} + \sigma_{yx}^\mathrm{V}) / \sigma_{xx}^\mathrm{N}] \simeq - \pi / 2$ for sufficiently low temperatures.
Note that the conductivity calculated from the TDGL expansion combined with the Hartree approximation [Eqs. (\ref{Eq:sigmaxx}) and (\ref{Eq:sigmayx})] can become worse toward the BEC side since the separation between $T^*$ and $T_\mathrm{c}$ is expected to be remarkable in the BEC side.
Nevertheless, we expect that $\mathit{\Theta}_\mathrm{H} \sim -\pi/2$ at low temperatures ($T\ll |\mu|$) where the dissipation is negligible. According to the Bogoliubov-de Gennes equations in the low-temperature limit, where only a few quantized levels exist in each vortex core \cite{Sensarma2006} and thus those states are hardly scattered by impurities and phonons. Further, spatially extended quasiparticles and collective modes have gapped spectra and thus the scattering between the vortex and thermally excited quasiparticles or collective modes are negligible. Consequently, the momentum transfer between the vortex and the background such as impurities, phonons and quasiparticles are negligible and vortex motion is similar to that in an ideal fluid, where a vortex flows with the velocity same as that of the fluid. In charged superconductors, the macroscopic vortex flow with averaged velocity $\bm{v}_\mathrm{v}$ induces spatially averaged electric field $\bm{E}=\bm{B}\times\bm{v}_\mathrm{v}$ with the spatially averaged magnetic field $\bm{B}$ \cite{Josephson1965}. When vortices flow parallel to the superflow, the transport current $\bm{J}_{\rm tr}$ and $\bm{v}_\mathrm{v}$ are anti-parallel and thus $\bm{E}$ and $\bm{J}_{\rm tr}\times\bm{B}$ are parallel, i.e. $\sigma_{xx}=0$ and $\sigma_{xy}<0$ for $\bm{B}$ parallel to $z$-axis. It then follows that $\mathit{\Theta}_\mathrm{H} = -\pi/2$.

\begin{figure}[h]
    \centering
    \includegraphics[scale=1.0]{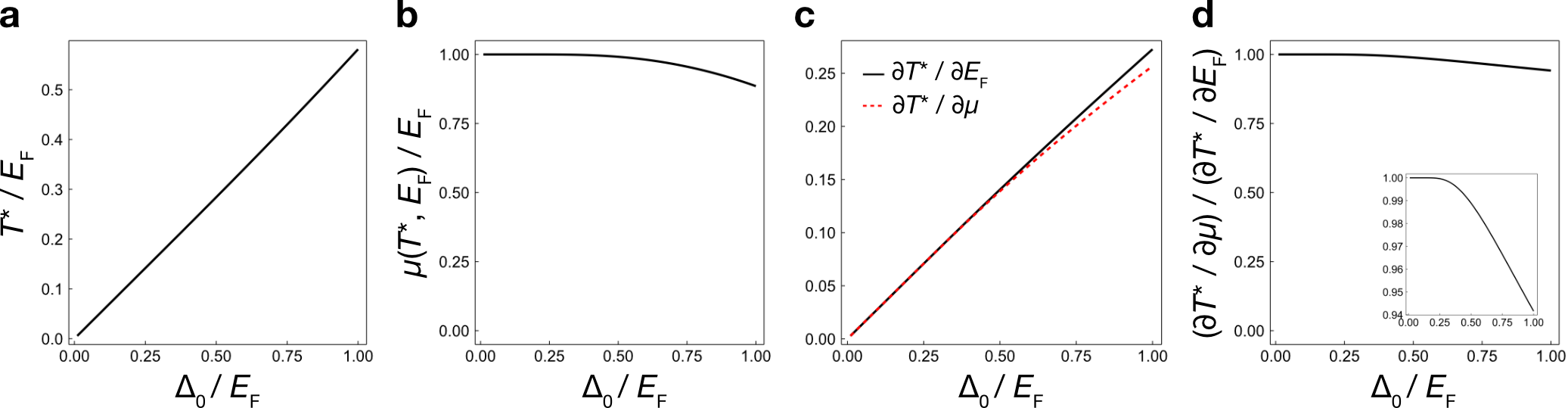}
    \caption{\textbf{Several quantities at the mean-field critical temperature.}\\
    (\textbf{a}) The mean-field critical temperature $T^*$ and (\textbf{b}) the chemical potential at $T^*$ as a function of the zero-temperature gap $\Delta_0$ divided by the Fermi energy $E_\mathrm{F}$.
    (\textbf{c}) $\partial T^* / \partial E_\mathrm{F}$ and $\partial T^* / \partial \mu |_{\mu = \mu (T^*, E_\mathrm{F})}$, as well as (\textbf{d}) the ratio of them, as a function of $\Delta_0 / E_\mathrm{F}$.
    For (\textbf{d}), an enlarged plot is shown in the inset.}
    \label{Fig:S08}
\end{figure}

Lastly, we discuss the difference between $\partial T^* / \partial \mu$ and $\partial T^* / \partial E_\mathrm{F}$ for the 2D Fermi gas model \eqref{Eq:Hamiltonian} within the mean-field approximation.
Using the two-particle binding energy $E_\mathrm{B}$ instead of the coupling constant $g$, we obtain the equation to determine $T^*$ for a given $\mu$~\cite{Botelho2006,Salasnich2013}:
\begin{equation}
    \int_0^\infty \mathrm{d} \varepsilon \left[ \frac{1}{2 \varepsilon + E_\mathrm{B}} - \frac{1}{2(\varepsilon - \mu)} \tanh \left( \frac{\varepsilon - \mu}{2 T^*} \right) \right] = 0.
    \label{Eq:mfeq}
\end{equation}
Note that $\Delta_0 =  \sqrt{2 E_\mathrm{B} E_\mathrm{F}}$ within the mean-field approximation~\cite{Salasnich2013}, where $\Delta_0$ is the superconducting gap amplitude at zero temperature.
For $\mu > 0$, we can rewrite Eq.~\eqref{Eq:mfeq} as~\cite{Gusynin1999}
\begin{equation}
    C_0 + \ln \frac{E_\mathrm{B}}{4 T^*} + \tanh \left( \frac{\mu}{2 T^*} \right) \ln \left( \frac{\mu}{2 T^*} \right) - \int_0^{\mu / 2 T^*} \mathrm{d} x \frac{\ln x}{(\cosh x)^2} = 0,
    \label{Eq:mfeq2}
\end{equation}
where $C_0 := - \int_0^\infty \mathrm{d} x \ln x / (\cosh x)^2 = 0.81878...$
Considering a small change in $\mu$ and the resulting change in $T^*$ in Eq.~\eqref{Eq:mfeq2}, we can obtain
\begin{equation}
    \frac{\partial T^*}{\partial \mu} = \frac{\tanh (\mu / 2T^*)}{1 + \tanh (\mu / 2T^*)} \frac{T^*}{\mu}.
    \label{Eq:derivative1}
\end{equation}
Then, we regard $T^*$ as a function of $E_\mathrm{F}$ ($= \pi \rho / m$), where $\rho$ is the particle density, and use the formula of $\mu$ for the 2D Fermi gas~\cite{Gusynin1999,Botelho2006,Salasnich2013}:
\begin{equation}
    \mu = T^* \ln (\mathrm{e}^{E_\mathrm{F} / T^*} - 1) =: \mu (T^*, E_\mathrm{F}).
    \label{Eq:mfeq_mu}
\end{equation}
Considering a small change in $E_\mathrm{F}$ and the resulting change in $T^*$ in Eqs.~\eqref{Eq:mfeq2} and \eqref{Eq:mfeq_mu}, we can obtain
\begin{equation}
    \frac{\partial T^*}{\partial E_\mathrm{F}} = \left[ \frac{E_\mathrm{F}}{T^*} + \left. \frac{\mathrm{e}^{(\mu - E_\mathrm{F}) / T^*}}{\tanh(\mu / 2T^*)} \frac{\mu}{T^*} \right|_{\mu = \mu (T^*, E_\mathrm{F})} \right]^{-1}.
    \label{Eq:derivative2}
\end{equation}
To summarize, we obtain $T^*$ and $\mu (T^*, E_\mathrm{F})$ by solving Eqs.~\eqref{Eq:mfeq2} and \eqref{Eq:mfeq_mu} simultaneously, and then we can determine $\partial T^* / \partial \mu |_{\mu = \mu (T^*, E_\mathrm{F})}$ and $\partial T^* / \partial E_\mathrm{F}$ from Eqs.~\eqref{Eq:derivative1} and \eqref{Eq:derivative2}, respectively.
In Supplementary Figs.~\ref{Fig:S08}(a-c), we show the calculated $T^*$, $\mu (T^*, E_\mathrm{F})$, $\partial T^* / \partial \mu |_{\mu = \mu (T^*, E_\mathrm{F})}$, and $\partial T^* / \partial E_\mathrm{F}$ as a function of $\Delta_0 / E_\mathrm{F}$.
We find that the ratio of $\partial T^* / \partial \mu |_{\mu = \mu (T^*, E_\mathrm{F})}$ to $\partial T^* / \partial E_\mathrm{F}$ is close to 1 even for moderate values of $\Delta_0 / E_\mathrm{F}$ [Supplementary Fig.~\ref{Fig:S08}(d)] such as $\Delta_0 / E_\mathrm{F} = 0.4$, which is a typical value observed in Li$_x$ZrNCl~\cite{Nakagawa2021} (Supplementary Fig.~\ref{Fig:S01}).

\bibliography{SI}
\bibliographystyle{apsrev4-2}


\end{document}